\RequirePackage[l2tabu]{nag}		
\documentclass[a4paper,11pt,leqno,oldfontcommands,  openany]{memoir} 

\usepackage{datetime}
\usepackage{ifpdf}
\ifdraftdoc 
	\usepackage{draftwatermark}				
	\SetWatermarkScale{0.3}
	\SetWatermarkText{\bf Draft: \today}
\fi
\newsubfloat{figure}
\newsubfloat{table}
\settrimmedsize{297mm}{210mm}{*}
\settypeblocksize{634pt}{448.13pt}{*} 
\setulmargins{4cm}{*}{*} 
\setlrmargins{*}{*}{1.5} 
\setmarginnotes{17pt}{51pt}{\onelineskip} 
\setheadfoot{\onelineskip}{2\onelineskip} 
\setheaderspaces{*}{2\onelineskip}{*} 
\checkandfixthelayout
\usepackage{fouriernc}
\usepackage[T1]{fontenc}
\OnehalfSpacing
\setsecnumdepth{subsection} 
\maxsecnumdepth{subsubsection}
\usepackage{calc,soul,fourier}
\makeatletter 
\newlength\dlf@normtxtw 
\newsavebox{\feline@chapter} 
\newcommand\feline@chapter@marker[1][4cm]{%
	\sbox\feline@chapter{%
		\resizebox{!}{#1}{\fboxsep=1pt%
			\colorbox{gray}{\color{white}\thechapter}%
		}}%
		\rotatebox{90}{%
			\resizebox{%
				\heightof{\usebox{\feline@chapter}}+\depthof{\usebox{\feline@chapter}}}%
			{!}{\scshape\so\@chapapp}}\quad%
		\raisebox{\depthof{\usebox{\feline@chapter}}}{\usebox{\feline@chapter}}%
} 
\newcommand\feline@chm[1][4cm]{%
	\sbox\feline@chapter{\feline@chapter@marker[#1]}%
	\makebox[0pt][c]{
		\makebox[1cm][r]{\usebox\feline@chapter}%
	}}
\makechapterstyle{daleifmodif}{

	\renewcommand\printchapternum{\null\hfill\feline@chm[2.5cm]\par}

} 
\makeatother 
\chapterstyle{daleifmodif}
\makepagestyle{myvf} 
\makeoddfoot{myvf}{}{\thepage}{} 
\makeevenfoot{myvf}{}{\thepage}{} 
\makeheadrule{myvf}{\textwidth}{\normalrulethickness} 
\makeevenhead{myvf}{\small\textsc{\leftmark}}{}{} 
\makeoddhead{myvf}{}{}{\small\textsc{\rightmark}}
\newcommand{\clearemptydoublepage}{\newpage{\thispagestyle{empty}\cleardoublepage}}
\makeindex
\usepackage{import}

\usepackage{lipsum}					
\usepackage{amsfonts} 					
\usepackage[centertags]{amsmath}			
\usepackage{stmaryrd}					
\usepackage{amssymb}					
\usepackage{amsthm}					
\usepackage{newlfont}					
\usepackage{layouts}					
\usepackage{graphicx}					
\usepackage{longtable,rotating}			
\usepackage[utf8]{inputenc}			
\usepackage{colortbl}					
\usepackage{wasysym}					
\usepackage{mathrsfs}					
\usepackage{float}						
\usepackage{verbatim}					
\usepackage{upgreek }					
\usepackage{latexsym}					

\usepackage{array}
\usepackage{enumitem}
\usepackage{booktabs} 
\usepackage{longtable}
\usepackage{pdflscape}
\usepackage{afterpage}
\graphicspath{{img/}}

\usepackage{listings, xcolor}

\definecolor{verylightgray}{rgb}{.97,.97,.97}

\lstdefinelanguage{Solidity}{
	keywords=[1]{anonymous, assembly, assert, balance, break, call, callcode, case, catch, class, constant, continue, constructor, contract, debugger, default, delegatecall, delete, do, else, emit, event, experimental, export, external, false, finally, for, function, gas, if, implements, import, in, indexed, instanceof, interface, internal, is, length, library, log0, log1, log2, log3, log4, memory, modifier, new, payable, pragma, private, protected, public, pure, push, require, return, returns, revert, selfdestruct, send, solidity, storage, struct, suicide, super, switch, then, this, throw, transfer, true, try, typeof, using, value, view, while, with, addmod, ecrecover, keccak256, mulmod, ripemd160, sha256, sha3}, 
	keywordstyle=[1]\color{blue}\bfseries,
	keywords=[2]{address, bool, byte, bytes, bytes1, bytes2, bytes3, bytes4, bytes5, bytes6, bytes7, bytes8, bytes9, bytes10, bytes11, bytes12, bytes13, bytes14, bytes15, bytes16, bytes17, bytes18, bytes19, bytes20, bytes21, bytes22, bytes23, bytes24, bytes25, bytes26, bytes27, bytes28, bytes29, bytes30, bytes31, bytes32, enum, int, int8, int16, int24, int32, int40, int48, int56, int64, int72, int80, int88, int96, int104, int112, int120, int128, int136, int144, int152, int160, int168, int176, int184, int192, int200, int208, int216, int224, int232, int240, int248, int256, mapping, string, uint, uint8, uint16, uint24, uint32, uint40, uint48, uint56, uint64, uint72, uint80, uint88, uint96, uint104, uint112, uint120, uint128, uint136, uint144, uint152, uint160, uint168, uint176, uint184, uint192, uint200, uint208, uint216, uint224, uint232, uint240, uint248, uint256, var, void, ether, finney, szabo, wei, days, hours, minutes, seconds, weeks, years},	
	keywordstyle=[2]\color{teal}\bfseries,
	keywords=[3]{block, blockhash, coinbase, difficulty, gaslimit, number, timestamp, msg, data, gas, sender, sig, value, now, tx, gasprice, origin},	
	keywordstyle=[3]\color{violet}\bfseries,
	identifierstyle=\color{black},
	sensitive=false,
	comment=[l]{//},
	morecomment=[s]{/*}{*/},
	commentstyle=\color{gray}\ttfamily,
	stringstyle=\color{red}\ttfamily,
	morestring=[b]',
	morestring=[b]"
}

\usepackage[utf8]{inputenc}
\usepackage[USenglish]{babel}
\usepackage[backend=biber,style=ieee,sorting=none,urldate=long]{biblatex}
\PassOptionsToPackage{hyphens}{url}
\usepackage{color}                    				
\usepackage[
	pdfauthor={James David Hackman},
	pdftitle={Forget-me-block: Exploring digital preservation strategies using DLT in the context of PIM},
	pdfsubject={MSc Computer Science thesis},
	pdfkeywords={Ethereum; Solidity; blockchain; Distributed Ledger Technology; digital preservation; Personal Information Management; calendar; smart contract},
	colorlinks=true,
	allcolors=black]{hyperref}              
\usepackage{url}						
\usepackage{memhfixc}					
\usepackage{enumerate}					
\usepackage{footnote}					
\usepackage{microtype}					
\usepackage{rotfloat}					
\usepackage{alltt}						
\usepackage[version=0.96]{pgf}			
\usepackage{tikz}						
\usepackage{arydshln}                   

\usepackage{tabularx}
\usepackage{multirow}
\usepackage{lscape}
\usepackage{makecell}
\usepackage{pdfpages} 

\usepackage{adjustbox}
\usepackage{geometry}

\usepackage{listings}
\usepackage{xcolor}

\definecolor{mygreen}{rgb}{0,0.6,0}
\definecolor{mygray}{rgb}{0.5,0.5,0.5}
\definecolor{mymauve}{rgb}{0.58,0,0.82}
 
\definecolor{darkgray}{rgb}{.4,.4,.4}
\definecolor{purple}{rgb}{0.65, 0.12, 0.82}
 
\lstdefinelanguage{JavaScript}{
keywords={typeof, new, true, false, catch, function, return, null, catch, switch, var, if, in, while, do, else, case, break},
keywordstyle=\color{blue}\bfseries,
ndkeywords={class, export, boolean, throw, implements, import, this, contract},
ndkeywordstyle=\color{darkgray}\bfseries,
identifierstyle=\color{black},
sensitive=false,
comment=[l]{//},
morecomment=[s]{/*}{*/},
commentstyle=\color{purple}\ttfamily,
stringstyle=\color{red}\ttfamily,
morestring=[b]',
morestring=[b]"
}
 
\usepackage{parskip}

\usepackage{array}
\newcolumntype{L}[1]{>{\raggedright\let\newline\\\arraybackslash\hspace{0pt}}m{#1}}
\newcolumntype{C}[1]{>{\centering\let\newline\\\arraybackslash\hspace{0pt}}m{#1}}
\newcolumntype{R}[1]{>{\raggedleft\let\newline\\\arraybackslash\hspace{0pt}}m{#1}}

\let\include\input

\usetikzlibrary{arrows,shapes,snakes,
		       automata,backgrounds,
		       petri,topaths}				
\newcommand{\pgftextcircled}[1]{                                                                    
    \setbox0=\hbox{#1}%
    \dimen0\wd0%
    \divide\dimen0 by 2%
    \begin{tikzpicture}[baseline=(a.base)]%
        \useasboundingbox (-\the\dimen0,0pt) rectangle (\the\dimen0,1pt);
        \node[circle,draw,outer sep=0pt,inner sep=0.1ex] (a) {#1};
    \end{tikzpicture}
}
\newcommand{\blackged}{\hfill$\blacksquare$}
\newcommand{\whiteged}{\hfill$\square$}
\newcounter{proofcount}

\let\oldsqrt\sqrt
\def\sqrt{\mathpalette\DHLhksqrt}
\def\DHLhksqrt#1#2{%
\setbox0=\hbox{$#1\oldsqrt{#2\,}$}\dimen0=\ht0
\advance\dimen0-0.2\ht0
\setbox2=\hbox{\vrule height\ht0 depth -\dimen0}%
{\box0\lower0.4pt\box2}}
\newcommand{\mycaption}[2][\@empty]{
	\captionnamefont{\scshape} 
	\changecaptionwidth
	\captionwidth{0.9\linewidth}
	\captiondelim{.\:} 
	\indentcaption{0.75cm}
	\captionstyle[\centering]{}
	\setlength{\belowcaptionskip}{10pt}
	\ifx \@empty#1 \caption{#2}\else \caption[#1]{#2}
}
\newcommand{\mysubcaption}[2][\@empty]{
	\subcaptionsize{\small}
	\hangsubcaption
	\subcaptionlabelfont{\rmfamily}
	\sidecapstyle{\raggedright}
	\setlength{\belowcaptionskip}{10pt}
	\ifx \@empty#1 \subcaption{#2}\else \subcaption[#1]{#2}
}
\usepackage{lettrine}
\newcommand{\initial}[1]{%
	\lettrine[lines=3,lhang=0.33,nindent=0em]{
		\color{gray}
     		{\textsc{#1}}}{}}
\theoremstyle{plain}

\theoremstyle{plain}

\theoremstyle{plain}
\theoremstyle{definition}

\theoremstyle{plain}

\theoremstyle{plain}

\theoremstyle{plain}

\definecolor{codegreen}{rgb}{0,0.6,0}
\definecolor{codegray}{rgb}{0.5,0.5,0.5}
\definecolor{codepurple}{rgb}{0.58,0,0.82}
\definecolor{backcolour}{rgb}{0.95,0.95,0.92}

\lstdefinestyle{mystyle}{
  backgroundcolor=\color{backcolour},   commentstyle=\color{codegreen},
  keywordstyle=\color{magenta},
  numberstyle=\tiny\color{codegray},
  stringstyle=\color{codepurple},
  basicstyle=\ttfamily\footnotesize,
  breakatwhitespace=false,         
  breaklines=true,                 
  captionpos=b,                    
  keepspaces=true,                 
  numbers=left,                    
  numbersep=5pt,                  
  showspaces=false,                
  showstringspaces=false,
  showtabs=false,                  
  tabsize=2
}

\begin{document}
%
%
%
%
%
\frontmatter
\pagenumbering{roman}
%
%
%
%
%
\begin{titlingpage}
\begin{SingleSpace}
\calccentering{\unitlength} 
\begin{adjustwidth*}{\unitlength}{-\unitlength}
\vspace*{13mm}
\begin{center}
\rule[0.5ex]{\linewidth}{2pt}\vspace*{-\baselineskip}\vspace*{3.2pt}
\rule[0.5ex]{\linewidth}{1pt}\\[\baselineskip]
{\HUGE Forget-me-block}\\[4mm]
{\Large \textit{Exploring digital preservation strategies using Distributed Ledger Technology in the context of personal information management}}\\
\rule[0.5ex]{\linewidth}{1pt}\vspace*{-\baselineskip}\vspace{3.2pt}
\rule[0.5ex]{\linewidth}{2pt}\\
\vspace{6.5mm}
{\large By}\\
\vspace{6.5mm}
{\large\textsc{James David Hackman}}\\
\vspace{11mm}
\includegraphics[scale=0.2]{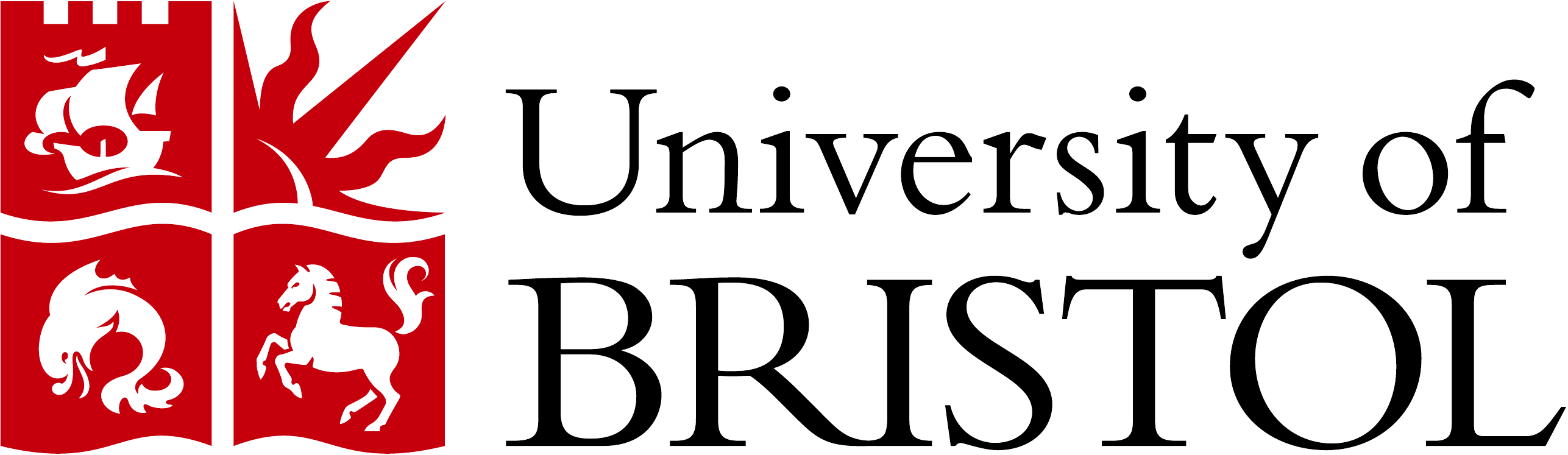}\\
\vspace{6mm}
{\large Department of Computer Science \\
\textsc{University of Bristol}}\\
\vspace{11mm}
\begin{minipage}{10cm}
\begin{center}
A dissertation submitted to the University of Bristol in accordance with the requirements of the degree of Master of Science by advanced study in Computer Science in the Faculty of Engineering.
\end{center}
\end{minipage}\\
\vspace{9mm}
{\large\textsc{15th September 2020}}
\vspace{12mm}
\end{center}
\begin{flushright}
\end{flushright}
\end{adjustwidth*}
\end{SingleSpace}
\end{titlingpage}

\newpage
%


\setcounter{page}{3}
%
%
%
\newgeometry{bottom=0.5cm} 
\chapter*{Executive Summary}
\begin{SingleSpace}
	\initial{R}eceived wisdom portrays digital records as guaranteeing perpetuity; as the New York Times wrote a decade ago: ``the web means the end of forgetting'' \cite{rosenWebMeansEnd2010}.  The reality however is that digital records suffer similar risks of access loss as the analogue versions they replaced - but through the mechanisms of software, hardware and organisational change.

The first two of these mechanisms are straightforward.  Software change relates to how data is encoded - for instance later versions of Microsoft Word often cannot access documents written with earlier versions \cite{sheesleyBackwardCompatibilityProblems2008}.  Likewise hardware formats obsolesce; even popular technologies such as the floppy disk reach a point where accessing data on these formats becomes increasingly difficult \cite{listRetrotechtacularFloppyDisk2019}.

	The third mechanism is however more abstract as it relates to societal structures, and ironically is often generated as a by-product of attempts to escape the first two risks.  In our efforts to rid ourselves of hardware and software change these risks are often delegated to specialised external parties.  Common use cases are those of conveying information to a future self, e.g. calendars, diaries, tasks, etc.  These applications, categorised as Personal Information Management (PIM) \cite[p.~12]{jonesFuturePersonalInformation2012}, negate the frailty of human memory.  Frequently these are outsourced at two removes - firstly by the individual to their employer (e.g. using a company system) and then by their employer to an external provider.  So enters organisational change risk; by the time the information is required the organisational chain that links user to data may be broken: the employer will have moved to a different provider, the employee will have left the company, the IS provider will be out of business or pivoted to new offerings.  

The advent of Distributed Ledger Technology (DLT) could help mitigate these risks; and has led to a re-evaluation of the relationship between data creation and ownership.  Although DLT is an imprecise term, it typically involves data storage across organisationally separate entities in a cryptographically secure form; and therefore could present a partial solution to the three risks outlined above.  This project makes a research contribution by:

\begin{itemize}[noitemsep,nolistsep] 
	\item Presenting the first research to date that applies DLT to the field of PIM and furthering design science state of the art by a novel implementation of a calendar application on the Ethereum blockchain.
	\item Extending current research in utilising DLT in digital preservation, namely by enacting a continuum approach within a DL that allows for transfer of ownership of digital objects as they transition from individual to collective relevance.
	\item Providing guidelines for future use of DLT within digital preservation; and demonstrating the utility of these guidelines via an evaluation of this project's software deliverables.
\end{itemize}
	In accordance with design science principles, post-submission, all source code will be made available on a public GitHub repository with an opensource licence, the software artefacts will be made live online and a blog post will be published linking code, live artefacts and implementation details \cite{hackmanForgetmeblockEthereumCalendar2020}.

\end{SingleSpace}

\restoregeometry 

%
%

\chapter*{Dedication and acknowledgements}
\begin{SingleSpace}
	I would like to thank my supervisor, Dr Ruzanna Chitchyan, for the generosity of her time.  I benefited tremendously from her expertise and approach.  Her challenge and interaction consistently added value and considerably improved the quality of this research project.  Thanks are also due to Lon Barfield of Simpleweb for taking the time to review my research proposal and provide useful feedback; and Andrew B Coathup (@abcoathup) of OpenZeppelin Forum for his guidance on smart contract implementation.\\

\end{SingleSpace}

\nopagebreak

%
%
%
%
%
%

\chapter*{Author's declaration}
\begin{SingleSpace}
\begin{quote}
I declare that the work in this dissertation was carried out in accordance with the requirements of the University's Regulations and Code of Practice for Taught Programmes and that it has not been submitted for any other academic award.  Except where indicated by specific reference in the text, this work is my own work. Work done in collaboration with, or with the assistance of others, is indicated as such. I have identified all material in this dissertation which is not my own work through appropriate referencing and acknowledgement. Where I have quoted or otherwise incorporated material which is the work of others, I have included the source in the references.  Any views expressed in the dissertation, other than referenced material, are those of the author.

\vspace{1.5cm}
\noindent
\textsc{J D Hackman}\\
\textsc{15th September 2020}\\
\href{mailto:j.hackman@preciouschicken.com}{j.hackman@preciouschicken.com}
\end{quote}

\end{SingleSpace}

\chapter*{Coronavirus impact statement}
\begin{SingleSpace}
	The closure of Bristol University study spaces and computer laboratories during this period, plus the inability to access the University of West England library under the SCONUL scheme; considerably impeded progress on this project.  Further details are contained within an extenuating circumstances form submitted to the Faculty.

\end{SingleSpace}

\newpage

\renewcommand{\contentsname}{Table of Contents}
\maxtocdepth{subsection}
\tableofcontents*
\addtocontents{toc}{\par\nobreak \mbox{}\hfill{\bf Page}\par\nobreak}
\cleardoublepage
%
\listoftables
\addtocontents{lot}{\par\nobreak\textbf{{\scshape Table} \hfill Page}\par\nobreak}
\newpage
\listoffigures
\addtocontents{lof}{\par\nobreak\textbf{{\scshape Figure} \hfill Page}\par\nobreak}
\clearemptydoublepage
%
%
\mainmatter
\let\textcircled=\pgftextcircled
\chapter{Introduction}\label{chap:intro}


\label{subsec:subsec01}

			\section{Distributed Ledger Technology}
			\label{sec:Overview}

			\initial{T}he genesis of Distributed Ledger Technology (DLT) was the pseudonymous Satoshi Nakamoto's Bitcoin white paper \cite{nakamotoBitcoinPeertoPeerElectronic2008} which introduced a form of peer to peer electronic cash that did not rely on intermediaries.  Despite emerging from obscure cryptoanarchist roots, rather than established academic or commercial circles, \cite{peckCryptoanarchistsAnswerCash2012}, Bitcoin successfully solved several key problems which had previously plagued private (i.e. non-state) digital cash projects \cite{7163021}.

\begin{figure}
				\includegraphics[width=0.9\columnwidth]{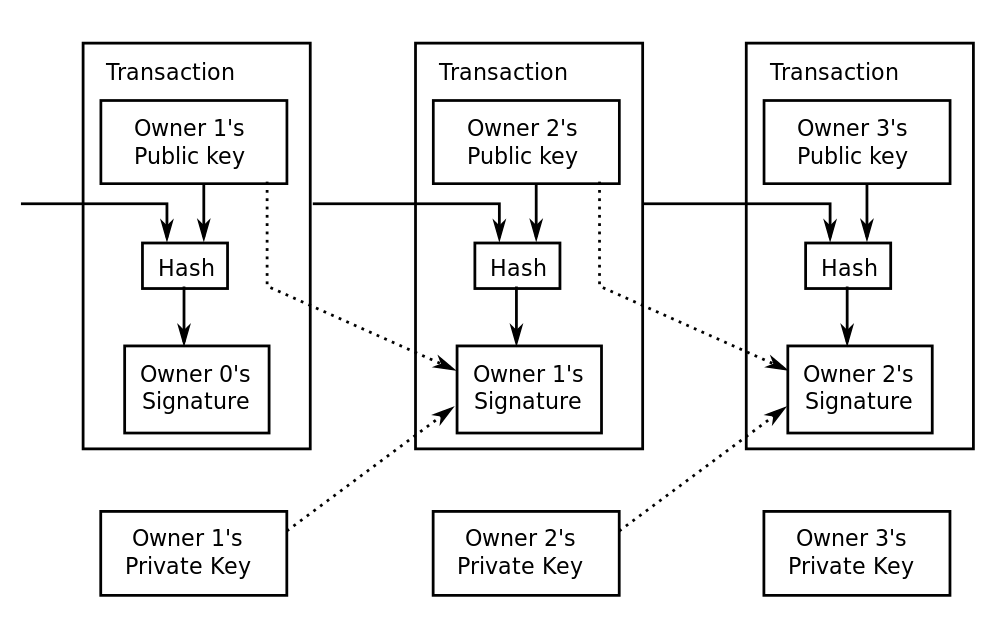}
				\caption[Bitcoin transactions]{Bitcoin transactions \cite{nakamotoBitcoinPeertoPeerElectronic2008}}\label{fig:BitcoinTransaction}
				\centering
			\end{figure}

			Bitcoin proved more resilient than previous efforts due to its skilful combination of economic and technical factors.  Transactions between users of the digital cash (e.g. Alice pays Bob one bitcoin) were cryptographically chained together into blocks (hence a blockchain -- see Figure \ref{fig:BitcoinTransaction}), and participants known as `miners' then competed to verify these blocks in a process known as `Proof of Work' -- a race to solve a computationally difficult puzzle.  The first miner to solve this puzzle was rewarded with an issue of bitcoin.  This economic incentive attracted a large number of actors who had more to gain in verifying the network and receiving bitcoin, than they did by attempting to falsify blocks (e.g. fake transactions so they could pay two people with the same bitcoin).  

			Although this innovation may have been over-hyped, some have for instance called it ``bigger than the internet'' \cite{carlsonStarSiliconValley2015}, it has been argued by Wenger \cite{Wenger2014} and others that it is a radically different way of organising information. Table \ref{table:Wenger2014} demonstrates how this technology is logically centralised (e.g. if Alice pays Bob one bitcoin, the network is agreed to that fact) but organisationally decentralised (i.e. not `owned' by any one enterprise, rather it is a standard that requires no permission to use or modify).
			
			\begin{table}
				\centering
				\begin{tabular}{ m{6em} m{7em} m{7em}  } 
					\toprule  
					& \begin{flushleft}Organisationally centralised\end{flushleft} & \begin{flushleft}Organisationally decentralised\end{flushleft} \\ 
						\midrule 
						\begin{flushleft}Logically centralised\end{flushleft} & e.g. Paypal & \textbf{***new***} blockchain \\
							\begin{flushleft}Logically decentralised\end{flushleft} & e.g. Excel & e.g. email \\ 
								\bottomrule
				\end{tabular}
				\caption[Foundational innovation of the blockchain]{Foundational innovation of the blockchain \cite{Wenger2014}}
				\label{table:Wenger2014}
			\end{table}

			Although not universally agreed \cite{ammousBlockchainTechnologyWhat2016}, many thought that this approach could be used to run other logically centralised but organisationally decentralised services apart from cash.  These efforts have been compared to ``a world computer'' \cite[p.~25]{antonopoulosMasteringEthereum2018} where services can be designed to run on platforms without seeking permission and once uploaded are both immutable and accessible by anyone \cite[p.~9]{grincalaitisMasteringEthereumImplement2019}.  This broader class of platforms have been defined as Distributed Ledger Technology (DLT) \cite{walportDistributedLedgerTechnology2015}, with the term Distributed Ledger (DL) implying that the field has grown to include structures others than blockchains, such as directed acyclic graphs \cite{bairdHederaPublicHashgraph2019}.  Although there is extensive debate on the conceptual definition of DLT \cite{osternBlockchainResearchDiscipline2019}, in its broadest terms it can be considered data storage across organisationally separate entities in a cryptographically secure form.

			Much of the excitement behind this innovation is due to the concept of a ``smart contract,'' a term predating DLT.  Here a contract, which is a written agreement between two parties for services or goods dependent on conditions being met, is evolved into a software based logic.  Szabo's original example is that of a vending machine - no human interaction is required: once a user inserts cash, the machine dispenses \cite{szaboFormalizingSecuringRelationships1997}.  This leads to an even more radical concept: Decentralised Autonomous Organisations, essentially whole enterprises that can achieve an objective, for instance trading commodities futures and making payouts to investors, with no human interaction \cite{krausBlockchainsSmartContracts2019}.

			This movement is motivated by a distrust of the current structures of authority; in the case of Bitcoin these are financial, but in wider DLT they are more generally corporate.  This is part of the ``Can't be evil'' movement \cite{aliCanBeEvil2017} where consumers of technology, frustrated by corporations such as Facebook mis-using their personal data \cite{cadwalladrRevealed50Million2018}, have decided to no longer take Google and kind at their word of ``Don't be evil.'' Instead they seek to enforce this in code using new decentralised platforms where no single entity has control, thereby limiting abuse on this massive scale.  User-facing software being built on these new platforms, often seeking to challenge these centralised incumbents, are known as Decentralised Applications (DApps) \cite{buterinEthereumWhitePaper2013}.  Some of the more radical imaginings even plan on rebuilding computer architecture from the ground up, starting with the operating system \cite{yarvinUrbitSolidStateInterpreter2016}.  

\begin{figure}
				\centering
				\includegraphics[width=0.9\columnwidth]{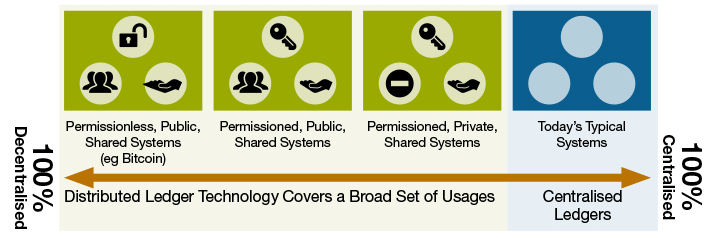}
				\caption[Walport's DLT variants]{Walport's DLT variants \cite{walportDistributedLedgerTechnology2015}}
				\label{fig:walport}
			\end{figure}

			Despite this strong cypherpunk element to DLT \cite{swartzWhatWasBitcoin2018}, it has also been co-opted by more established forces.  Although the original DLs were intended as open to everyone (e.g. `permissionless'), this was followed by the rise of `permissioned' DLs, often founded by networks of corporates who saw an advantage to pooling data, but wished to retain control of the system.  As Walport \cite{walportDistributedLedgerTechnology2015} discusses this in itself is a sliding scale, with DLs having a variety of degrees of openness and transparency (see Figure \ref{fig:walport}).

\section{Digital preservation}

			This technology though also has benefits for strategies promoting digital preservation, defined as ``the ability to sustain the accessibility, understandability and usability of digital objects in the distant future regardless of changes in technologies and in the [communities] that use these digital objects'' \cite{rabinovici-cohenSIRFSelfcontainedInformation2011}.  Data that exists on these DLs, or ``world computers'' \cite[p.~25]{antonopoulosMasteringEthereum2018}, in an immutable and accessible form is in theory no longer affected by the fortunes of one corporate entity, and is spread amongst many different hardware instances.  This distributed system for applications and data therefore has utility against potential threats to digital preservation (see further discussion at Section \ref{sec:Threat}).

The UK Joint Information Systems Committee states digital preservation is not a ``stand alone activity'' but rather should be embedded in how institutions ``manage and approach digital information and resources on an ongoing basis'' and recommends a lifecycle approach \cite{beagrieDevelopmentDigitalPreservation2005}.  Given that this research project seeks to implement software artefacts, digital preservation will be primarily approached via the service design phase \cite[p.~81]{itsmfinternationalITILFoundationHandbook2012} of the lifecycle.  Thus the focus is on designing and building software for long term use -- where long term is defined by the Open Archival Information System as ``long enough to be concerned with the impacts of changing technologies, including support for new media and data formats or with a changing user community'' \cite{theconsultativecommitteeforspacedatasystemsReferenceModelOpen2012}.

\section{Personal Information Management}
\label{sec:pim}

The present project seeks to examine one aspect of digital preservation: Personal Information Management (PIM).  Jones provides perhaps the fullest description of what PIM in practice is:

\begin{quotation}
``the activities a person performs in order to acquire or create, store, organize, maintain, retrieve, use and distribute the information needed to meet life’s many goals ... PIM places special emphasis on the organization and maintenance of personal information collections in which information items, such as paper documents, electronic documents, email messages, web references, handwritten notes, etc., are stored for later use and repeated re-use.'' \cite[p.~5]{jonesKeepingFoundThings2008}
\end{quotation}

This activity is a central feature of many information systems, but there also exist applications created specifically for this: namely calendars, task managers, address books, etc \cite[p.~60]{jonesFuturePersonalInformation2012} which are a mainstay of our working lives.  Considering their centrality to the modern workplace (see Chapter \ref{chap:Context}), this research will consider digital preservation in the context of these particular applications.

			\section{Aims}

			The aims of this research project divide into objectives, the knowledge and steps required to complete the project; and deliverables, artefacts which may be of wider use following completion of the project.

			Given the potential breadth of this topic, this project utilises stretch targets, which may not be achievable within the given timescale.  Objectives and deliverables have therefore been broken into necessary, desirable and advanced as indicated below.

			\subsection{Objectives}\label{sec:Objectives}
			\subsubsection*{Necessary}
			\label{sec:ObjectivesNecessary}
			\begin{enumerate}
				\item Gain personal experience of designing and implementing DApps.
				\item \label{list:ConsiderDL} Consider types of DLs; contrast, compare and decide on most suitable for DApp implementation.
				\item Understand mechanisms chosen DL uses for storage and retrieval of data. 
				\item Design a DApp that seeks to establish independence from the three pillars of threat to digital preservation: software, hardware and organisational change.
			\end{enumerate}
			\subsubsection*{Desirable}
			\label{sec:ObjectivesDesirable}
			\begin{enumerate}[resume]
				\item Using experiences gained from DApp implementation, understand the challenges of designing systems that focus on digital preservation, mapping that to real world context that users will experience over the course of their lives.

				\item Understand the wider challenges implicit in using DLs for data storage, especially those relating to personal information management.
			\end{enumerate}
			\subsubsection*{Advanced}\label{sec:ObjectivesAdvanced}
			\begin{enumerate}[resume]
				\item Drawing on academic literature, practitioners and example scenarios; synthesise themes of threat and mitigation in digital preservation as applied to DLT.
			\end{enumerate}

			\subsection{Deliverables}
			\label{sec:Deliverables}
			\subsubsection*{Necessary}
			\label{sec:DeliverablesNecessary}
			\begin{enumerate}
				\item A smart contract that provides notification of previously recorded information, at a predetermined future date, or prompted by a retrieval request.

				\item A DApp based around the smart contract above where the user is notified  via an interface such as a web browser.
			\end{enumerate}

			\subsubsection*{Desirable}
			\label{sec:DeliverablesDesirable}
			\begin{enumerate}[resume]
				\item  Guidelines, based on reflection and related literature, aimed at developers and researchers designing DApps to address digital preservation.
			\end{enumerate}

			\subsubsection*{Advanced}
			\label{sec:DeliverablesAdvanced}
			\begin{enumerate}[resume]
				\item Extending the feature base of the DApp so incorporating some functionalities of a Personal Information Management application.
				\item Empirically validating the guidelines at Deliverable 3 and DApp at Deliverable 4 against analysis of example scenarios.  
			\end{enumerate}

			\section{Research contribution}\label{sec:IntroContribution}

			Completion of the Objectives and Deliverables will make a research contribution in the following way:

\begin{itemize} 
	\item Presenting the first research to date that applies DLT to the field of PIM and furthering design science state of the art by a novel implementation of a calendar application on the Ethereum blockchain.
	\item Extending current research in utilising DLT in digital preservation, namely by enacting a continuum approach within a DL that allows for transfer of ownership of digital objects as they transition from individual to collective relevance.
	\item Providing guidelines for future use of DLT within digital preservation; and demonstrating the utility of these guidelines via an evaluation of this project's software deliverables.
\end{itemize}
	In accordance with design science principles, post-submission, all source code will be made available on a public GitHub repository with an opensource licence, the software artefacts will be made live online and a blog post will be published linking code, live artefacts and implementation details \cite{hackmanForgetmeblockEthereumCalendar2020}.

			\section{Architecture}\label{sec:IntroSummary}
			DLT is still nascent, meaning the task of designing DApps is more challenging than building the equivalent on traditional centralised services.  Indeed even the choice of which of the myriad DL to build on is bewildering, as will be explored in Chapter \ref{chap:LitReview}.

\begin{figure}
				\centering
				\includegraphics[width=0.9\columnwidth]{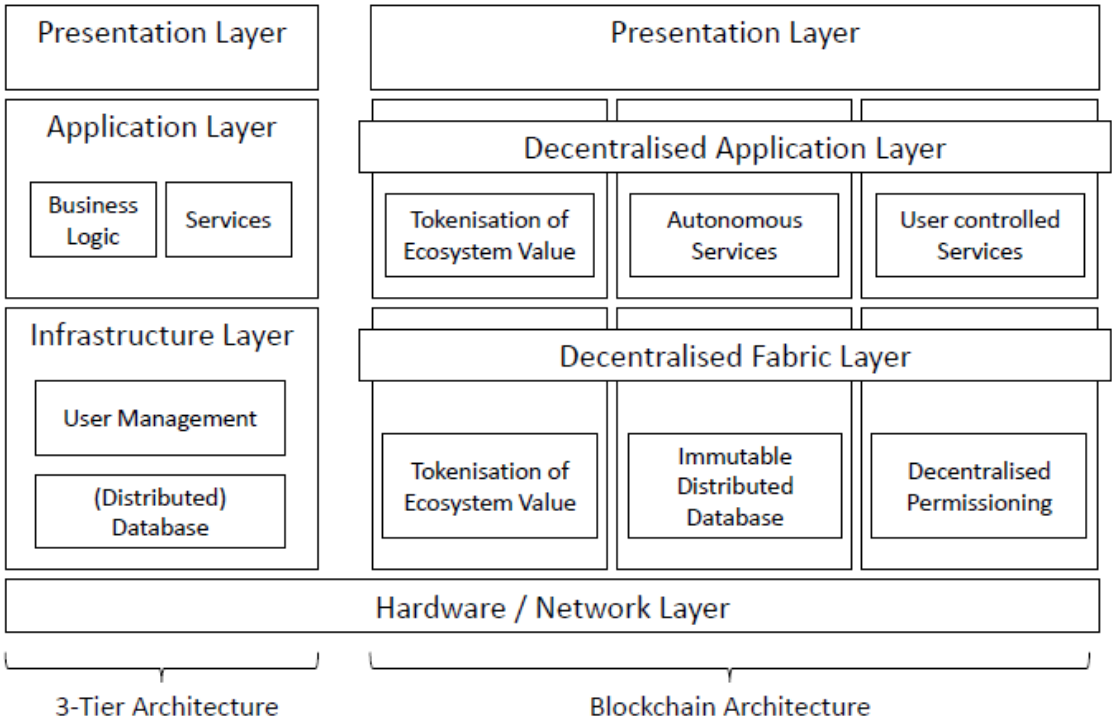}
				\caption[Glaser's three-tier blockchain architecture]{Glaser's three-tier blockchain architecture \cite{glaserPervasiveDecentralisationDigital2017}}
				\label{fig:three-tier}
			\end{figure}

			This is not an entirely brave new world however, and touchpoints exist with non-DL systems.  One for instance is the architectural view: DApps built at the present time can still be seen through the lens of a traditional three layer architecture \cite[p.~24]{sumathi2007fundamentals}, which Glaser \cite{glaserPervasiveDecentralisationDigital2017} visualises at Figure \ref{fig:three-tier}. 

To put this in more concrete terms, this project sees DApp design implementing this three-tier architecture as follows:

			\begin{itemize}
				\item \textbf{Presentation or Client}.  The element that the user interacts with.  This might take the form of a Web browser, a mobile phone `app' or a desktop application e.g. an email client.  Due to the novelty of DLT many of these clients do not have a native ability to interact with a DL (though there are exceptions such as the Brave web browser \cite{bondyRoadBrave2019}) and have to use extensions such as Metamask \cite{leeMetaMaskMonthlyFebruary2020}, which connects web browsers to the Ethereum DL.
				\item \textbf{Application}.  This is the DL itself and contains the business logic (i.e. smart contract) required by the application.  An example of this might be RSK \cite{lernerRSKBitcoinPowered2019} -- this is a DL which works in concert with the Bitcoin blockchain. 
				\item \textbf{Storage or Data}.  DLs theoretically can store not only business logic, but also persistent data as a more standard database might.  However DLs such as RSK are typically not used for this, but rather concentrate on the compute function and store data elsewhere.  Clearly if this `elsewhere' is on a centralised cloud provider then the benefits of a decentralised model are lost.  This challenge is now starting to be addressed by decentralised storage options such as IPFS \cite{benetIPFSConentAddressed2014} or BigchainDB \cite{bigchaindbBigchainDBBlockchainDatabase2018} which provide solutions for DLs through a variety of mechanisms, but tend to rely on financially incentivising a wide variety of participants to store other people's (encrypted) data.  This project will refer to these entities collectively as decentralised databases (DDb).
			\end{itemize}

\begin{figure}
				\centering
				\includegraphics[width=0.9\columnwidth]{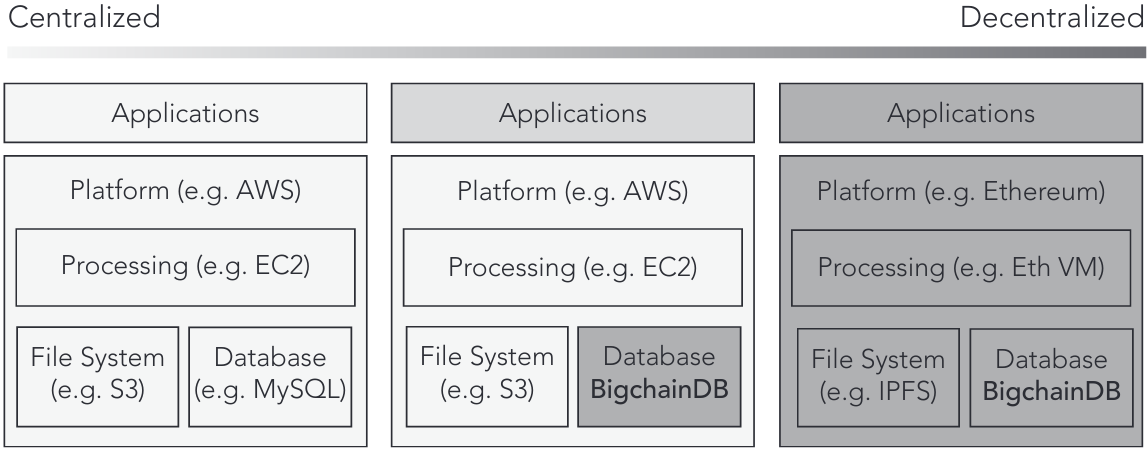}
				\caption[BigchainDB's centralised to decentralised architectural stack]{BigchainDB's centralised to decentralised architectural stack \cite{bigchaindbBigchainDBBlockchainDatabase2018}}
				\label{fig:bigchaindb}
			\end{figure}

In reality these three layers are blurred: the original blockchain, Bitcoin, has little ability to handle smart contracts, and stores all data `on chain' -- even if that data, bitcoin transactions, is very specific.  Nor is there consensus -- Glaser's \cite{glaserPervasiveDecentralisationDigital2017} definition differs from this project as his third layer, which he calls the `fabric layer', is intended as ``the execution environment   for   smart   contract   languages.''  Xu et al \cite{xuTaxonomyBlockchainBasedSystems2017} meanwhile in their taxonomy would define this third layer as ``off-chain item data.''  This lack of agreement is not surprising: as Ostern \cite{osternBlockchainResearchDiscipline2019} identifies there is considerable ``conceptual fuzziness'' within the research community as to what a DL even is, so naturally architectural models will also not cohere.  BigchainDB, a DDb \cite{bigchaindbBigchainDBBlockchainDatabase2018}, also demonstrate at Figure \ref{fig:bigchaindb} that transition from centralised to decentralised is not necessarily binary either.  Regardless of this lack of definition, a three-tier architecture remains a useful model for the purposes of this research.

Given that the presentation layer, a web browser or desktop application, is comparatively well understood, research effort will instead concentrate on exploring the choices that exist within the application or DL layer.  The software deliverables however will implement a presentation layer -- ultimately smart contracts that are only accessible via a command line interface are unlikely to offer many insights into digital preservation strategies that extend into real life applications.  This project will also seek to use a DL as a storage layer, rather than a specific decentralised storage solution.  Although this is a far from optimum solution (see Section~\ref{sec:EvalStorage}), it takes into account both time constraints and the aims of this project to produce design science artefacts (see Chapter~\ref{chap:Methodology}) rather than production-ready software.  Before any deliverables can be produced, it is necessary to examine the threats to digital preservation this project seeks to mitigate and potential methods to select a DL to build this architecture on.

\let\textcircled=\pgftextcircled
\chapter{Background}
\label{chap:LitReview}

			\initial{T}his chapter will first examine the problem space that this application explores, or `the threat'; focus will then turn to determining the most appropriate DLT platform to address this problem (partly fulfilling Objectives 2 and 5 at Section \ref{sec:Objectives}).  Due to the layered complexity of DLT \cite{glaserPervasiveDecentralisationDigital2017} and its conceptual vagueness  \cite{osternBlockchainResearchDiscipline2019}, the field remains immature and is characterised by a lack of clarity \cite{deshpandeDistributedLedgerTechnologies2017}.  Given this landscape therefore selecting an appropriate architecture is not a trivial problem and one which would benefit from an examination of the literature.

			\section{The threat}\label{sec:Threat}

			This research project looks towards DL as a potential solution, or at least mitigation, of the risks individuals and organisations face in trying to ensure digital preservation.  Ross states that the risks to data loss can be categorised as: human error, computer virus, natural disaster, hardware or software problems \cite[p.~118]{rossApproachingDigitalPreservation2006}.  Recognising that this is a considerable topic in itself, and putting aside human error, this research will collapse these factors into three strands: threats from hardware, software and organisational change.

			An explanation of all these three terms, and how in reality they tend to intertwine is, convincingly given by Vines et al \cite{vinesAvailabilityResearchData2014}.  Here data from 516 biological studies published between 1991 and 2011 was analysed.  Of those studies only 121 (23\%) had data sets that were still accessible, with the authors calculating the odds of a data set being extant falling by 17\% per year.  Responses to why this data was no longer available included the data being on inaccessible hardware (e.g. stolen), or on zip or floppy disks which authors no longer had appropriate hardware to access.  Further data sets were beyond access because the email addresses provided with many of the sample papers were no longer active, and therefore the original authors could not be contacted.  This draws into sharp focus the organisational change aspect of data loss.  Academics move between institutions; but the email addresses appended to publications do not.  Likewise organisational change was given as a major factor in Markwell and Brooks study of web page ``link rot'' in biochemistry and molecular biology resources \cite{markwellLinkRotLimits2003}.  Here 20\% of URLs over the course of a two year period were no longer accessible: both due to internal re-organisation and extra-organisational domain name transfer. 

			Looking at settings other than research, Huvila et al explore several digital preservation case studies featuring both hardware and software issues \cite{huvilaContinuumThinkingContexts2014}.  One case study finds sound artists losing confidence in gaining access to their music collections due to their storage on now obsolete MiniDiscs.  They also find `software' change issues at Uppsala University Library where archivists find WordPerfect files difficult to catalogue due to file type recognition problems.  Indeed Bird and Simons \cite{birdSevenDimensionsPortability2002} in their authoritative account of the problems of storing linguistic data, note that records that are over 100 years old are often easier to access for researchers than those stored on modern media.  
			
			Indeed digital preservation is a common challenge: any knowledge worker will be familiar with the difficulties of sharing document formats \cite{kosekOfficeDocumentFormat2008}.  Likewise the concept of hardware obsolescence is familiar in the marketplace \cite{nahmEffectCompatibilitySoftware2001} -- as Breen et al noted the newer the technology the ``shorter the life expectancy'' \cite{breenTaskForceEstablish2003}.  

			Although one might imagine the market not promoting hardware longevity, it is also problematic for professionals whose core concern is digital preservation.  In 1983 a BBC project sought to recreate the famous Domesday book of 1082 and recruiting academic and technical skills at a national level, asked over 14,000 schools and community groups to provide a snapshot of British life through images, text and video which would be saved in digital format \cite{tapperDomesdayProjectEducational1986}.  Unfortunately, unlike the original, it was unreadable within 15 years, not a 1000 years, due to hardware obsolescence issues \cite{mckieDigitalDomesdayBook2002}.  Consequently the material was considered of such importance a rescue effort was launched to safeguard the data by the National Archives \cite{darlingtonDomesdayReduxRescue2003} -- but clearly this amount of effort is not sustainable for all data.

			Meanwhile the threat from organisational change is, if anything, accelerating.  The literature suggests recent decades have seen a move from `career jobs' to a more precarious labour market \cites{jacobyAreCareerJobs1999}[p~30-33]{ngCareerConcepts2018}, across both the blue \cite[p.~67-68]{pupoWalmartizationMcJobJobs2010} and white collar \cite{styhreNewFormsProfessional2017} spectrum. Nor it is only the relation of the employee to the firm that is changing; the role of knowledge management within the organisation, has shifted to greater emphasis on social tools and a collapse of personal and professional boundaries \cite{archer-brownHybridSocialMedia2018} resulting in ``blogs, wikis, social bookmarking, data mashups, editing platforms, or media sharing'' \cite{vonkroghHowDoesSocial2012} forming a wider part of the mix -- of course all of this spanning various corporate boundaries (not a new phenomena in itself \cite{mckemmishEvidenceMe1996}).  Carr \cite{carrITDoesnMatter2003} writing two decades ago heralded a time when organisations would buy in information services as a commodity like electricity; this has come to pass: organisations routinely switch contracts from Microsoft's Office 365 to Google's G-Suite to back again.  So even when an employee stays with the same company, there is no guarantee that the data landscape beneath their feet is a solid one.

			Turning to the literature on digital preservation, all three risk factors can be found, including that of organisational change  \cite{rosenzweigScarcityAbundancePreserving2003, beagrieDigitalCurationScience2008}.  The field is extensive and multi-disciplinary, unsurprising given that these risks cover the gamut of hard to soft sciences.  Related fields include computer science, librarianship, archival science, records management, digital forensics, museum curation and the management of information systems to name a subset \cite{leeWhereArchivistDigital2011}.  Indeed occasionally the term ``digital curation'' is used for this collective, which although typically referring to digital research data (which would cover the example of Vines et al \cite{vinesAvailabilityResearchData2014} above), can also be seen to include the processes needed for best practice data creation and management \cite{beagrieDigitalCurationCentre2004} and indeed touches on collections of personal information \cite{beagrieDigitalCurationScience2008}.  This project will use the term digital preservation however, primarily as the use of this term appears more focused on system design \cite{leeStateArtPractice2002, gladneyTrustworthy100yearDigital2005, westerlundDesigningAutomatedDigital2020}; although it is acknowledged there is considerable overlap.  It is also something of an irony that the field of digital preservation itself is also at risk from its nemesis:  CAMiLEON \cite{wheatleyMigrationCAMiLEONDiscussion2001} was a UK / US research project focused on migrating the BBC Domesday project above -- the homepage detailing their efforts is now only available via the internet archive \cite{finneyBBCDomesdayProject2015}.

Within the digital preservation literature the continuum model \cite{upwardModellingContinuumParadigm2000} shares similarities with the concerns of this research project.  The continuum model, although designed as a framework to archival activities, is relevant as it looks at the creation of data via a spacetime continuum.  Looking further than a lifecycle approach to records (potentially analogous in IS development to the Waterfall method), it considers concentric rings around an object as the utility of that record ripples from owner, to the organisation (and relevant intra-organisation groupings), to wider society.  To return to Vines et al's datasets \cite{vinesAvailabilityResearchData2014} we can see that disruption, due to the three threats (software, hardware, organisation) during that outwards transition, can prevent full realisation of the knowledge contained within the data.  Elsewhere the continuum approach is defined as the ``holistic attention to the full span of design, creation, management, use, and reuse of digital objects'' \cite{leeWhereArchivistDigital2011} -- this all encompassing framework suggests that this research, concerned with the design of applications that safeguard digital preservation, fits within its remit.

The continuum model also firmly covers the space of PIM as demonstrated by Huvila et al \cite{huvilaContinuumThinkingContexts2014}.  One example they provide is how a personal letter from an individual to a public authority, when placed in a public collection, becomes part of a larger narrative about public engagement at that point in history.  Continuum theory encompasses far wider subjects than those covered by this project however: for instance it considers transformative topics such as participatory research on the inclusion of social and cultural matters into record keeping \cite{e8152c89fe4b42eab05a6557406ea841}.

			Indeed this research project is not the first to see a link between DLT and continuum.  Lemieux et al \cite{lemieuxBlockchainTechnologyRecordkeeping2019} posit that blockchain systems have an affinity with the continuum model, given immutable records are fixed in their own continuum of a chain of blocks.  Nor is the continuum theory's link with DLT purely theoretical -- researchers have cited it whilst attempting practical implementations. Project Archangel \cite{collomosseARCHANGELTrustedArchives2018} have experimented with using the Ethereum blockchain for preservation of digital records.  Concerned that public trust in the institutions responsible for archiving may decline, and consequently public faith in the integrity of the records they store; they looked to DLT for solutions.  Hashes of documents stored (and metadata such as file type), were uploaded to the Ethereum blockchain, and then made accessible via a web front end -- essentially meaning that anyone querying the provenance of a document could ensure its integrity without relying on the institution itself.  Interestingly this approach makes a nod to all three pillars of the change threat:

			\begin{itemize}
				\item Organisation: The hash of the record is being stored in a way that is accessible independently of the organisation holding the item.
				\item Software: Data regarding the application required to access the data (as catalogued by the UK National Archives' DROID (Digital Record Object IDentification) \cite{uknationalarchivesDROIDUserGuide2020}) is being stored with the record. 
				\item Hardware: Being stored on Ethereum the record is duplicated across many different machines in many different parts of the globe.
			\end{itemize}

			Although this is hardly a complete solution -- the records themselves are still subject to loss within the institutions responsible for their preservation; it is a good example of a practical experiment to overcome the risks or organisational, software and hardware change.  

			Archangel is not the only project in this space: InterPares Trust Chain being a similar example \cite{bralicModelLongtermPreservation2017a}.  Garcia-Barriocanal, Sanchez-Alonso and Sicilia \cite{garcia-barriocanalDeployingMetadataBlockchain2017a} also implement a proof of concept for storing metadata on Ethereum, but take this further by discussing using a third-layer storage option (i.e. IPFS) to store the digital objects themselves.  They still however see the processing ability as laying outwith the DL and using a form of migration or emulation.

			Lemieux \cite{lemieuxTypologyBlockchainRecordkeeping2017} attempts to categorise the intersections of DLT with record-keeping, suggesting a model that groups systems as types of ``mirror,'' ``digital record'' or ``tokenized.''  This encompasses a range of options, from where institutions simply replicate current mechanisms but using DLT (e.g. ``mirror'') to where the records stand in for the objects themselves -- ``tokenized.''  She foresees potential paradigm shifts in how digital preservation is considered. 

Woodall and Ringel however highlight some of the tensions that underlie these efforts to include DLT in digital preservation \cite{woodallBlockchainArchivalDiscourse2019}.  In the same way that DLT has been considered conceptually fuzzy \cite{osternBlockchainResearchDiscipline2019}, they point to the range of discourses that are brought to the subject depending on the different backgrounds of the participants.  They identify a contradiction between those professionals with responsibilities for digital preservation who would rather see trust reside in institutions that can ``protect rights and mediate disputes;'' and DLT proponents who see those same professionals as a ``single point of failure through centralized control'' who should be replaced.  The scope of this debate is outwith this project, although it is hoped that realistic limitations of this project's software artefacts, and by extension DLT, is presented in Section \ref{sec:DiscEval}.

It is also the case that not all agree on the size of the threat: Kirschenbaum \cite[p.~14]{kirschenbaumMechanismsNewMedia2008} seeks to `correct' assumptions about the vulnerability of digital objects; pointing to how magnetic media is frequently recoverable under the right conditions.  This does appear however a minority view, the majority of the literature emphasises the threat rather than edge cases.  This project therefore posits that threats to digital preservation from software, hardware and organisational risks are real and warrant solutions.  Chapter \ref{chap:Results} will examine how we might go about designing DApps to counter some of these threats.

			\section{Choosing a Distributed Ledger solution}\label{sec:ChoosingDL}

			Although there is considerable literature on whether an enterprise should use DLT to solve a problem or not \cite{peckBlockchainWorldYou2017, wustYouNeedBlockchain2018, BlockchainHype2018, betzwieserDecisionModelImplementation2019}; for those who have already decided that a DL is the right option, there is less guidance on selecting an appropriate DL amongst the many options.  

			The stand out exception however is Farshidi et al \cite{farshidiDecisionSupportBlockchain2020}, who offers a comprehensive online tool to help developers choose an appropriate DL.  This Decision Support System (DSS) attempts to map features of DLs, such as zero-knowledge proofs, alongside metrics taken from an ISO software quality model.  Despite the strengths of this approach there are clearly shortcomings -- a primary method for gathering data on DLs is reading white-papers and interviews with DL subject matter experts.  Unfortunately both of these approaches have their disadvantages: white-papers tend to be overly optimistic about a DL's capability and are a poor substitute for hands on development experience.  Likewise selecting `experts' in this nascent field is problematic -- the first blockchain is only just over a decade old, and for part of that period was unknown outside of a small community of cryptographers \cite{Maurer2013}, meaning there are few with real domain experience.

			Where the literature is not focusing on specific implementations, much of the remaining literature does not seek to provide guidance on DL selection but rather serves to classify DL thematically; e.g. technical criteria \cite{tasca2017ontology, ozdemirAssessmentBlockchainApplications2019} or sectors where it might be applied \cite{yli-huumoWhereCurrentResearch2016, risiusBlockchainResearchFramework2017, labazovaHypeRealityTaxonomy2019, schedlbauerBlockchainDigitalCurrencies2018, akramAdoptionBlockchainTechnology2020, guptaBlockchainResearchScientometric2020}.

			This lack of guidance in the literature is problematic given the amount of technologies that self-describe as blockchain or DL.  
			One cryptocurrency website lists 7,053 coins \cite{CoinGeckoCryptocurrencyPrices2020}, although not a measure of blockchains (as one DL can have multiple coins or tokens), it does hint at the size of the problem.  Even Farshidi et al's far more sober decision support tool \cite{farshidiDecisionSupportBlockchain2020} lists 29 technologies -- which is a sizeable number of alternatives.

			One answer to this problem might emerge not from the field of computer science, but economics.  Hayek argues a rational economic system cannot be chosen by committee, as no one body can hold all the different variables that the market is guided by \cite{hayek1945use}.  Rather price acts as indicator of where resources should be allocated.  Although the objective here is not to choose an economic order, this approach might be useful -- what metric driven by individual participants might show the utility of a DL?  Although market capitalisation of a cryptocurrency could be one such factor (indeed it is for Farshidi et al \cite{farshidiDecisionSupportBlockchain2020}) given that this is often driven by speculation rather than utility \cite[p.~132-4]{goodman_2019}, this seems a poor choice.  A suitable metric therefore might be what DL developers actually spend their time developing for; StackOverflow analysis being one method of devising this already established in the CS literature \cite{baruaWhatAreDevelopers2014a, 6693332, mayGenderDifferencesParticipation2019}.

			This project will therefore take a blended approach to DL selection: using Farshidi et al's tool to first narrow down blockchain choice, followed by real-world analysis of data taken from StackOverflow.  

	\newgeometry{bottom=22mm} 
			\begin{landscape}
				\begin{table}
					\centering
		\renewcommand{\arraystretch}{2}
					\begin{tabular}{ m{14em} m{10em} m{35em} } 
						\toprule
						Feature & MoSCoW & Description \\
						\midrule
						Application layer & Must-Have & The blockchain has a layer on which (decentralized) applications can be developed and run \\
						Smart-contracts & Must-Have & Programmed contracts that are enforced by computer protocols which enable transactions -- in this domain on a blockchain \\
						Turing Complete & Must-Have & The Virtual Machine that is used by the blockchain platform is Turing Complete\\
						Programming Language Support & Must-Have & The blockchain supports at least one programming language to create new applications / contracts / tokens / blockchains \\
						Resilience Technologies & Should-Have & This blockchain utilizes at least one (including niche) mainly resilience related feature\\
						Technology Maturity & Should-Have & The maturity of a blockchain platform relative to other platforms \\
						Transaction Speed & Could-Have & The transaction speed of a blockchain platform relative to other platforms\\
						\bottomrule
					\end{tabular}
					\caption{Features chosen from Farshidi et al's DSS \cite{farshidiDecisionSupportBlockchain2020}}
					\label{table:DSSCriteriaTerms}
				\end{table}
			\end{landscape}
	\restoregeometry

			\begin{landscape}
				\begin{table}
					\centering
		
					\renewcommand{\arraystretch}{1.2}
					\begin{tabular}{ m{9em} r m{35em}  } 
						\toprule  
						DL & \makecell{ StackOverflow \\ questions }  & Search term  \\ 
						\midrule 
						Ethereum & 34,705 & ethereum or ethereum.stackexchange.com \\ 
						Hyperledger & 10,517 & hyperledger\\ 
						EOS & 9,068 & eos or eosio.stackexchange.com \\ 
						R3 Corda & 4,045 & corda\\ 
						NEO & 1,195 & ``neo'' -neo4j -``neo 4j''\\ 
						Cosmos Network & 569 & cosmos -azure -microsoft -[fiware] -[cosmos] -[azure] -[azure-cosmosdb] \mbox{-[fiware-cosmos]} \\ 
						JP Morgan Quorum & 67 & [quorum]\\ 
						QTUM & 48 & qtum \\ 
						Wanchain & 6 & wanchain \\ 
						Chain & Not known & Not applicable\\ 
						\bottomrule
					\end{tabular}
					\caption[Farshidi et al's DSS \cite{farshidiDecisionSupportBlockchain2020} results compared to StackOverflow questions]{Farshidi et al's DSS \cite{farshidiDecisionSupportBlockchain2020} results compared to StackOverflow questions as of 14 April 2020 (latter author's own work)}
					\label{table:DSSResult}
				\end{table}

				\begin{table}
				\centering
				\begin{tabular}{ m{10em} r } 
					\toprule  
					Classification & Count \\ 
					\midrule 
					Task or Todo & 17 \\ 
					Calendar & 2 \\ 
					Contact Manager & 2 \\ 
					\bottomrule
				\end{tabular}
				\caption[Competitor analysis of DApps]{Competitor analysis of DApps as of 25 May 2020 (author's own work)}
				\label{table:CompetitorDApps}
			\end{table}
			\end{landscape}

			Table \ref{table:DSSCriteriaTerms} shows the seven features chosen from the DSS.  Of a total of 75 features available, only these were considered critical.  MoSCoW prioritisation (Must have, Should have, Could have and Will not have) was applied to the seven features, grouped as:

			\begin{itemize}
				\item Must have.  Relate to the project deliverable of a DApp -- this cannot be built without an \textit{application layer} with \textit{programming language support} in a \textit{Turing complete} manner.  
				\item Should have.  Relate to the digital preservation aspect -- designing a DApp that will be resistant to data loss you wish to guard against the three pillars of change (hardware, software, organisation) -- \textit{resilience} guards against software loss, \textit{maturity} against organisational change.
				\item Could have.   This was assigned to \textit{transaction speed} -- a calendar or contact app would need some form of relatively quick response, but it is difficult to place too much emphasis on this without more understanding of the practicalities of how this would work.
			\end{itemize}
			
			This feature set resulted in a DSS response as at Table \ref{table:DSSResult}, all of which scored a 100\% match.  This list of DL was then taken and analysed against the numbers of questions asked on StackOverflow, the results of which were added to Table \ref{table:DSSResult}.  

			Table \ref{table:DSSResult} also shows the search term used in StackOverflow, which requires further explanation.  A number of DLs had their own separate StackOverflow site in which case the total questions from this site were added to the number of questions on the primary site, where this occurred the top level domain name is shown as well as the search term.  

			Additionally several of the search terms were highly ambiguous.  NEO for instance shares its namesake with a NoSQL database, while Cosmos also refers to a toolkit for building operating systems.  Indeed the official StackOverflow tag for Cosmos refers to the toolkit and not the DL.  In both instances non-related terms were excluded from the search (indicated with a minus at Table \ref{table:DSSResult}).  Quorum is also used in common language, therefore just instances of tagged questions were included (tags are represented by square brackets at Table \ref{table:DSSResult}).  The DSS also proposed ``Chain'' as a suitable DL, however the link provided from the DSS was broken and with no further information plus such an ambiguous name it was excluded, although it had appeared in the literature previously \cite{8048839}.

			This analysis shows a clear leader -- Ethereum.  It is also the case that another DL, Quorum, is simply a permissioned fork of Ethereum.  The top three DLs; so Ethereum, Hyperledger and EOS were therefore short-listed as contenders.  As the case of ``Chain'' shows however this field moves quickly, and in the course of this literature review DLs were encountered which were potentially not known to Farshidi et al \cite{farshidiDecisionSupportBlockchain2020}, yet were worthy of further consideration.   Therefore added to this list was Hedera Hashgraph \cite{bairdHederaPublicHashgraph2019} which shows promise due to its novel approach of using a directed acyclic graph rather than a blockchain to order data, so resulting in faster transactions; and Tezos which offers formal verification of contracts \cite{goodmanTezosSelfamendingCryptoledger2014}.  Competitor analysis also revealed another potential platform to shortlist -- Blockstack, the main differentiating feature of which is that it hopes to act as a ``a fullstack for decentralized computing'' \cite{muneebBlockstackDecentralizedComputing2019}. This is particularly interesting as it is attempting to seamlessly integrate the application and storage layer as referred to at Section \ref{sec:IntroSummary}.  Having selected a list of potential DLs to build upon, through study of the literature and ecosystem analysis, a decision will be taken at Chapter \ref{chap:Implementation}.

\section{Competitor analysis}
\subsection{Commercial}

			In the course of researching DLs, competitor analysis was conducted to determine what other PIM applications exist within the DLT space.  \textit{State of the DApps} \cite{stateofthedappsStateDAppsExplore2020},  which is a cross-DL listing website for DApps featured in previous research \cite{wuFirstLookBlockchainbased2019}, was searched using the tags `tasks' and `calendar' (there was no `contact' tag); and then by browsing the entire `media' and `social' categories for similar DApps.  Searching \textit{State of the DApps} revealed that a large number of DApps used the Blockstack DL, so their DApp store was also analysed by browsing through all DApps available.  A list of the total number of related DApps found is at Table \ref{table:CompetitorDApps}.

			Table \ref{table:CompetitorDApps} is if anything over-representative however: 28.5\% (6 in total) of the DApps listed were not functional: they either were simply a link to a GitHub repository (with no instantiated application), did not load or did not operate (e.g. loaded but the user interface did not function).  Of those that did work a considerable number appeared to be in a beta stage of development and may not have been suitable for long term use.

			For comparison a search for `calendar' on the Google Play store returned 250 applications.  Although for time reasons this list was not analysed further; considering that the Android operating system represents just one choice for users, it puts into perspective the early stage of development of DLT.

			It also points to the conclusion that applications built using this technology, although not entirely novel, are still exploring relatively uncharted territory; so demonstrating the added value of this research.

\subsection{Research}

\begin{figure}
				\includegraphics[width=0.9\columnwidth]{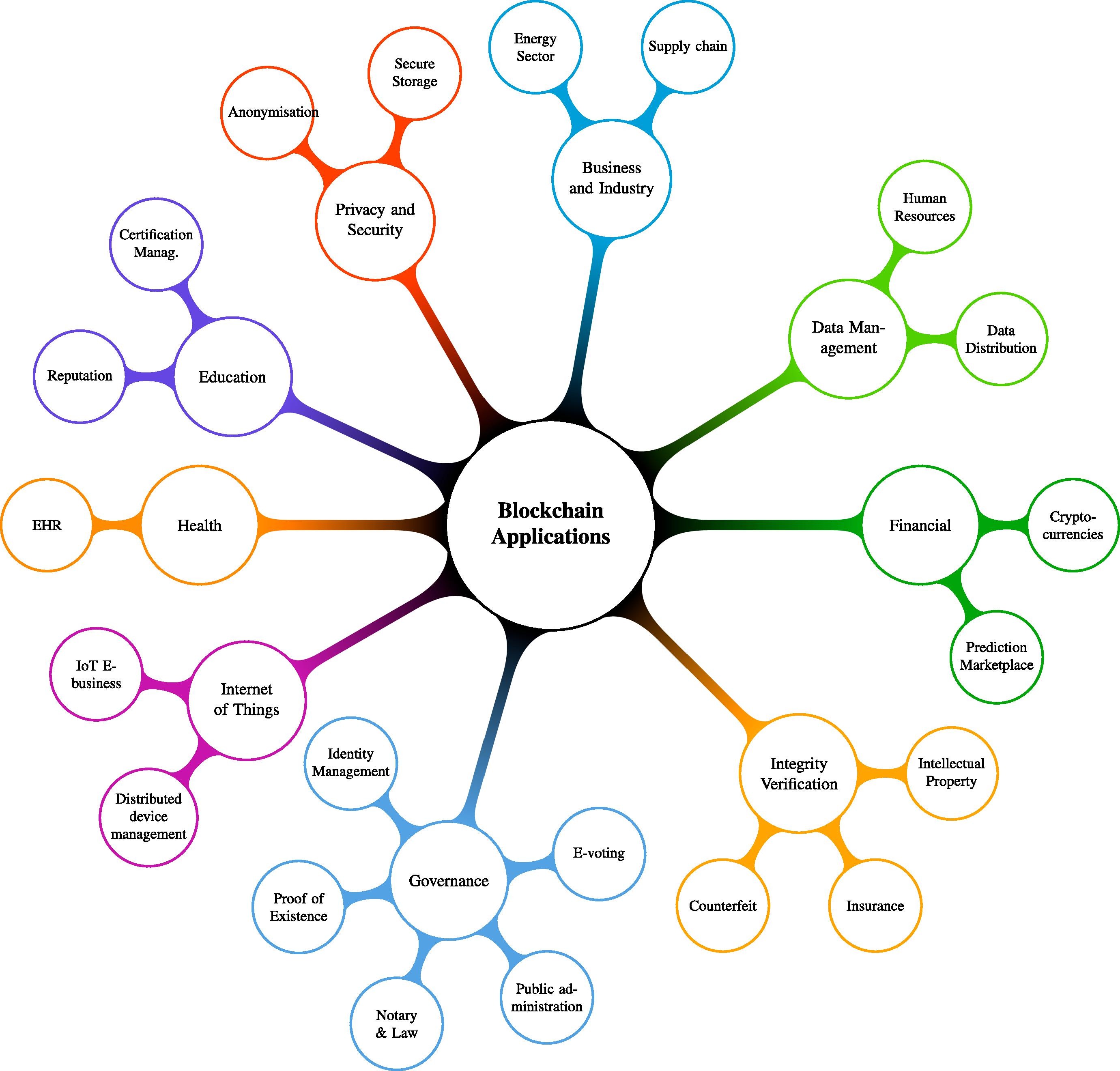}
				\caption[Systematic literature review mindmap of blockchain applications]{Systematic literature review mindmap of blockchain applications \cite{casinoSystematicLiteratureReview2019}}\label{fig:mindmap}
				\centering
			\end{figure}

Figure \ref{fig:mindmap} shows DApps that have been implemented in the course of research as determined by Casino et al's recent systematic literature review~\cite{casinoSystematicLiteratureReview2019}.  Noticeably they do not find any evidence of PIM applications, as defined at Section \ref{sec:pim}, within their review.  The closest category is data management, which is primarily concerned with storing data on a DL and covers several authors mentioned in Section \ref{sec:Threat}.

From title alone Zichichi et al \cite{zichichiEfficiencyDecentralizedFile2020} may qualify as similar work as they consider DLT within a ``Personal Information Management System'' (PIMS).  The authors do not define PIMS, but rather cite a report from an EU agency \cite{europeandataprotectionsupervisorOpinion2016EDPS2016} which defines it as ``new technologies and ecosystems which aim to empower individuals to control the collection and sharing of their personal data'' and goes on to state alternative terms are ``personal data stores, personal dataspaces or personal data vaults.''  Indeed the example of PIMS the study uses is that of geo-tracking mobile phones on a public transport system -- which does not fit within the definition at Section \ref{sec:pim}.  PIM consists of the ``activities'' performed by an individual to manage their information \cite[p.~5]{jonesKeepingFoundThings2008} -- clearly with geo-tracking a mobile phone there is no attempt at individual curation; it occurs regardless of the user.  So although location can play an important part in PIM (for instance recording where a photo was taken), the research by Zichichi et al \cite{zichichiEfficiencyDecentralizedFile2020} is considered distinct from the field of PIM researched in this project.

This project therefore presents the first research implementation of DLT for PIM.

\section{Summary}

This section has looked at the threat posed by hardware, software and organisational threats to digital preservation and concluded that the risk is very much extant.  Next the technological landscape of DLT was considered and specific platforms selected based on relevant criteria.  Lastly it was considered what other PIM implementations exist within DLT, both in the commercial and academic space -- this analysis concludes this project will be presenting a novel implementation given the current state of the art.

With this background covered, and the need for effective digital preservation established, the project will next consider exactly how these challenges of digital preservation are actualised over the course of a professional career.

\let\textcircled=\pgftextcircled
\chapter{Context}\label{chap:Context}

\initial{T}he preceding examination of the literature establishes that digital preservation, especially in the realm of PIM, is a difficult task.  This chapter will seek to link how these challenges to digital preservation might play out over the course of a typical career or life.  This contributes towards Objective 5 (see Section \ref{sec:Objectives}) --  namely understanding the real world context of digital preservation.

Two industry sectors were selected to highlight digital preservation over the course of a career: Engineering and Health.  For each scenario the literature was analysed to form a high level view of the forms of challenge to digital preservation that might occur over a longitudinal period.  This analysis was then overlaid over the career of a prototypical professional within these sectors.  Resources such as LinkedIn \cite{linkedinWelcomeYourProfessional2020} and the National Careers Service \cite{nationalcareersserviceCareersAdviceJob2020} were used to ensure accuracy was obtained and that the scenarios chosen were true to life.

\section{Scenario 1: Alice the Marine Engineer}

This scenario was primarily based on the findings of Pikas \cite{pikasPersonalInformationManagement2007}, who conducted qualitative research into the PIM strategies of a number of senior engineers.  This detailed study draws into focus the real life PIM requirements of engineers and relates it well to the challenges they face in meeting these needs with information technology.  The principal downside of the study is its limited number of participants, although that was partly mitigated by the interviewees' decades' worth of combined experience.

One datapoint that emerges from this research is the importance of personal contacts, stored partly via contact book software; the participants used this information to rely ``on networks of expert colleagues as information sources.''  Not only were the participants interested in who they were speaking to, they also wanted to map when these conversations occurred, with one engineer stating they ``couldn't keep track of all the phone calls I was making, who I was talking to, and who was calling me.''  After recording these interactions the interviewee notes these ``became collection points and then I had a calendar or a time sequence'' -- clearly illustrating that agendas and calendars were also an important part of their practice.  The participants also noted that this networked expertise was not confined to those working within their organisation or geographically close to them, but rather was something they assembled over the course of their career. 

With a high level overview in place, now let us imagine the challenges facing a particular individual following this career path: Alice.  The individual might start her career by completing a MEng at the University of Sheffield in Mechanical Engineering.  This period sees them initiating their network of fellow engineers in the form of their classmates and potentially academics.  Following their successful graduation they apply for a commission in the Royal Navy and join as a Marine Engineer.  The next sixteen years sees them actively deployed as a Marine Engineering Officer onboard frigates and destroyers, interspersed with periods working in support roles ashore, for instance providing technical advice to those charged with legal accountability for the safe operation of equipment.  During this period she extends her network both with fellow Service personnel, and with those in the wider industry such as Babcock and BAE engineers.  Following the arrival of a young family she decides she wishes to spend less time operationally deployed, leaves the Royal Navy and joins a Bath based consultancy as a Maritime Safety Consultant.  Unfortunately during the next economic downturn this consultancy fails, and she joins the MoD Civil Service working for a project team that oversees upkeep periods for minor warships.  A number of years after leaving the Navy, and having established herself in the civilian world, Alice rejoins the Naval Reserve where she mentors young engineers, so developing a new passion for education. Coming to the end of her working life and retiring from the Civil Service, she takes up a voluntary position on the Board of Governors at a local STEM Academy where she intends to pass on some of her skills to the next generation.

What digital preservation issues will be faced over the course of this long and varied career?  Given the importance of capturing a network of experience as identified by Pikas \cite{pikasPersonalInformationManagement2007}, this will be difficult, given the current state of software.  The systems she first uses at University will almost certainly cease on her finishing her academic studies.  It is likely that she will have to use informal means to keep in contact with her classmates (e.g. social media), however the marketplace is likely to see provider shifts over a period of this length (e.g. Facebook replacing MySpace \cite{boyd2013white}).  Over the course of her Naval career as she switches between poorly connected operational deployments to shore based roles she will move between different systems and into different positions, meaning PIM continuity breaks.  Additionally given enough time it is likely the RN will undergo information system transformations (e.g. moving from MS Office 365 to Google G-Suite or to in-house solutions) with resulting disruption.  Leaving the RN and joining the civilian world will bring an end to all the digital material accrued on Naval systems -- and it is unlikely that for security reasons it can be exported.  The failure of the Bath Consultancy she joins will almost certainly see her digital history at that company deleted.  Within the MoD Civil Service and Naval Reserves, even if the same systems as the Regular service are used, it will be unlikely that previously recorded information can be retrieved.  With her last position on the Board of Governors, much of her work is completed on her own hardware, using MS Office 365 -- but this time associated with her `sch.gov.uk' account credentials, and therefore inaccessible to any of her other collected information.

Clearly this variety of systems is going to present a challenge to any career-long PIM strategy.  Classmates from university are unlikely to be easily contacted in her Bath consultancy role; and as a school governor she might remember receiving a pertinent brief at a conference during her time in the Naval Service, but would find it difficult to ascertain when or by whom it was delivered to look up details.  It is clear that corporate systems cannot be relied upon to maintain the user's PIM continuity over her career, which means that if she wishes to do this she will need to take matters into her own hands and rely on her own private resources.  This brings its own problems though -- generally corporations will take responsibility for migration as they switch from one hardware or file format to another; if you take personal responsibility then this becomes another task for the user to handle.  It is unlikely that she will be using the same hardware as a governor as when she was an undergraduate -- she therefore has to transfer key data and keep on top of changing data formats over the course of her life if she expects to access the personal information she curated as an undergraduate by the time she is a school governor.  It is almost certain that the PIM strategy outlined by Pikas \cite{pikasPersonalInformationManagement2007} is, although not impossible to maintain, going to find the current centralised software landscape an obstacle.  These findings tally with a study of engineers' PIM by McAlpine, Hicks and Tiryakioglu \cite{mcalpineDigitalDivideInvestigating2011} who find that engineers primary complaint is the ``non-linkage of different technologies or it systems'' which is reflected both in software (i.e. the inability to share outside of specific applications) and hardware (i.e. having to migrate files manually from one system to another).

\section{Scenario 2: Bob the Clinical Psychologist}
\label{sec:clinpsych}

This second scenario will consider the use of PIM for a professional within a healthcare setting -- specifically mental health.  Although the literature does not feature equivalently comprehensive studies as Pikas \cite{pikasPersonalInformationManagement2007}, primarily as far more research concerns Health Information Management (HIM), there are still a number of useful studies on PIM within medical settings.  Both Molazadeh et al \cite{molazadehInvestigatingUsePersonal2017} and Devitt and Murphy \cite{devittSurveyInformationManagement2004} find widespread use of PIM within medical universities and acute hospital trusts respectively.  Sedghi et al \cite{sedghiQualitativeStudyPersonal2015} also find widespread PIM within a medical university faculty, with a key finding being an extensive cross-over between using personal and professional systems for storing information -- mostly as participants had low levels of trust in their employers' networks capacity for digital preservation.  Ironically however relying on their own systems did result in data loss and difficulty in recalling where items were stored.  Focusing on PIM tools specifically Payne, Wharrad and Watts \cite{payneSmartphoneMedicalRelated2012} find junior doctors rely extensively on calendar applications, similarly Barrett, Strayer and Schubart \cite{barrettAssessingMedicalResidents2004} note that residents' most popular mobile applications, after medical reference books, were PIM tools such as calendars, to-do lists and address books.  They also highlight that a key concern for these clinicians were catastrophic data loss and security (especially with regards to patient confidentiality).  

Much of this literature relates to the immediate concerns in dealing with patients.  However longitudinal research has also been conducted.  Marques-Sanches et al \cite{marques-sanchezImportanceExternalContacts2018} conducted social network analysis of contacts versus job performance in a health care setting, finding that external ties contribute to improving the performance of physicians.  Interestingly this result was found for physicians, but not nurses; potentially because physicians were responsible for more decision making processes and therefore needed to rely on inter-professional networks to enable knowledge translation.  The study also suggested the importance in healthcare of personal contacts, and hence the flow of knowledge, between academic and clinical settings.  Consistent with this, Tasselli \cite{tasselliSocialNetworksProfessionals2014} shows networks have implications for the spread of innovation through healthcare settings, while Meltzer et al \cite{meltzerExploringUseSocial2010} suggests external networks are key to information dissemination; although systematic reviews suggest more research is required before patient care outcomes can be categorically linked \cite{sabotUseSocialNetwork2017}.  Research by Ward et al \cite{wardRoleInformalNetworks2014} suggests similar improvements for middle-managers in the sector -- suggesting that networks are used to ``maintain knowledge collectively''; and finding that relationships were informal, fluid and long lasting as managers moved between public, private and voluntary associations.  One particular aspect where networks might affect healthcare practitioners more than those in other occupations is self-care following traumatic events: Wald \cite{waldOptimizingResilienceWellbeing2020} notes that resilience is strengthened by contact with a network of health care colleagues.  A strong body of research therefore suggests that networks are important to those in a healthcare setting; it seems likely that digital preservation of this information, via PIM, throughout their careers is valuable.

Having given a high level overview of why PIM is professionally important within the healthcare sector, let us turn to the possible career path of one such professional -- in this case a Clinical Psychologist called Bob.  His initial degree is at Sheffield University where he reads for a BSc in Psychology.  During his course he becomes interested in mental health and on graduation joins an NHS primary care service (IAPT - Improving Access to Psychological Therapies) in Sheffield as a Psychological Wellbeing Practitioner, primarily delivering cognitive behavioural therapy.  Wishing to advance his career he gains a job as an Assistant Psychologist at the Blankshire General Hospital before applying for a Doctorate in Clinical Psychology at Southampton University.  A successful graduation from this course sees him specialising in Clinical Health Psychology and working at a number of hospitals in the South.  During this period he also takes on a role as regional co-ordinator for the British Psychological Society.  Wanting to develop his research interests Bob secures a position at UCL where he joins the Institute of Epidemiology \& Health Care investigating the links between mental wellbeing and diabetes.  Having worked closely with the charity Diabetes UK on various funded projects and appreciating the work they do, he decides to join them in their advocacy department.  Frequent involvement with parliamentary committees expands his network within governmental circles, and leads to a role as an adviser in the Department of Health.  Eventually he reaches the end of his working career and takes on a non-remunerated position as an Honorary Research Fellow at Imperial College's Faculty of Medicine.

Clearly many issues that affect the Engineer are going to be just as relevant for the Psychologist.  His career sees a constant shift between sectors: health, academia and government in this case.  This will mean a constant move between hardware and software, making it difficult to cohere PIM.  Much of this will be provided by his employer, but not exclusively - in his co-ordination role for the British Psychological society he is likely to organise events and activities using social media such as WhatsApp and Facebook.  These bring their own challenges, especially in the migration of data from these platforms. Both the Engineer and Psychologist will be concerned about security of their data but in different contexts: with the former it will be operational security concerns, the latter patient confidentiality. 

\section{Summary}

Many of the current issues with PIM explored above are known to the field.  Lush's literature review \cite{lushFundamentalPersonalInformation2014} highlights a major theme being fragmented information -- particularly the fact that the data the user requires is split across distinct applications.  The same review also reiterates the problems already discussed as to how software incompatibility and hardware obsolescence can cause longer term challenges to PIM.  This is not an idle concern: research suggested that better PIM strategies lead to better professional outcomes for knowledge workers \cite{hwangPersonalInformationManagement2015}.  Next will be considered this project's approach to examining this problem.

\let\textcircled=\pgftextcircled
\chapter{Methodology \& Method}\label{chap:Methodology}

			\initial{W}hile noting Hassani's concerns that ``computing research is not supported by well-defined, globally accepted methods'' \cite{hassaniResearchMethodsComputer2016} this project nonetheless adopts a research paradigm of radical humanism \cite[p.~80]{johnsonUnderstandingManagementResearch2000}. This allies a subjectivist perspective combined with the `sociology of radical change' -- an  uncommon approach within the information systems community \cite{organInformationSystemsRisk2013}, although previously explored within DL literature \cite{HomVir20191}.  Subjective due to the realisation that DL is portrayed as potentially ground breaking and over hyped at the same time, primarily as it promises a break from the socially constructed realities that much of IS mirrors.  At the same time radical change is present in the deliverables illustrating how alternative technological realities might be realised.  Correspondingly the research's philosophical position is one of critical realism -- the IT stack that currently exists; reliant on centralised, corporate offerings from AWS, etc; is a reflection of dominant economic and social structures, and not necessarily of technological constraints.  Nakamoto did not invent any truly new technology, but rather combined existing components in a seminal way \cite{narayananBitcoinAcademicPedigree2017}.  It is therefore not too great a jump to imagine in a world where power structures were differently formulated, a Nakamoto like intervention may have happened earlier, and we might now have a technological perspective that solves problems in radically different ways.  

Within this position the approach will be deductive -- that is research questions will be answered from an understanding of the literature.  A mono-method qualitative data study will be conducted -- DApp deliverables will be evaluated against guidelines and scenarios for their contribution to digital preservation \cite{fridgen2018developing}; a subjective process but one that will enable a comparison to be conducted.  The project will take a cross-sectional time horizon, that is the deliverables will be concerned with the state of DL at the present time.  The final piece of the research methodology are techniques and procedures which can be defined as implementation-based -- the researcher realises software artefacts to stand in for a generic class of IS solution~\cite{barghResearchSkillsSoftware2014a}.

			The combination of a radical humanism paradigm with an implementation-based procedure will necessarily result in deliverables that do not represent ``best of breed'' solutions.  Rather they will represent another way of looking at technology and how society chooses to use IS tools to address problems.  In this project the problem will be that of digital preservation, and the alternative will be DLT in contrast to the dominant centralised model.

\section{Design science research}

Glaser \cite{glaserPervasiveDecentralisationDigital2017} posits that examples of design science research are particularly useful when exploring DLT -- due to the novelty and complexity of the technology practical instantiations are a useful way to advance knowledge; a call recognised by others in DLT research \cite{derossiComprehensiveBlockchainArchitecture2019}.

There is not however firm agreement on what design science research, or simply design science, is -- leading Baskerville to deduce that it is easier to state categorically what it is not \cite{baskervilleWhatDesignScience2008}.  Typically it does involve designing IT artefacts (often running code) that seeks to further understanding by solving a problem \cite{vaishnaviDesignScienceResearch2019}.  This project will follow a design cycle of iteration with the primary artefacts produced being software artefacts (DApps) \cite{hevnerThreeCycleView2007}.  Also delivered will be guidelines as to designing DApps which give regard to digital preservation; and an evaluation of the DApp software artefacts against the guidelines designed.  Furthermore this project will pay due regard to the seven design science guidelines laid out by Hevner et al \cite{hevnerDesignScienceInformation2004}:

			\begin{enumerate}
					\label{list:designscience} 
				\item \textbf{Design as Artefact}.  Will produce viable artefacts in form of instantiation (DApp) and model (guidelines).
				\item \textbf{Problem relevance}.  Digital preservation is a `wicked' problem \cite{vaishnaviDesignScienceResearch2019} as validated by research (see Section \ref{sec:Threat}).
				\item \textbf{Design evaluation}.  Artefacts rigorously evaluated using `Descriptive -- Scenario' method to demonstrate utility \cite{hevnerDesignScienceInformation2004}.
				\item \textbf{Research contributions}.  As outlined at Section~\ref{sec:IntroContribution}, the artefacts are the research contribution, applying a new model of technology (and wider organisation) to a societal problem.
				\item \textbf{Research rigour}.  Architectural choices made following analysis of relevant research (see Section~\ref{sec:ChoosingDL}), artefacts analysed for integrity at Section~\ref{sec:EvalIntegrity}.
				\item \textbf{Search process}.  Search process based on design cycle with artefact becoming more valuable and relevant through each iteration (see Chapter \ref{chap:Implementation}).
				\item \textbf{Research communication}.  Research will be made publicly accessible (see Section~\ref{sec:IntroContribution}) and will attempt to communicate at a level commensurate with the field.
			\end{enumerate}

			\section{Phases of implementation}

			With the theoretical underpinnings of this approach covered, the remainder of this chapter will look at the practical steps required to implement this work.  

			\subsection{Smart contract: Deliverable 1}

			Following DL selection the first task will be to design and deploy a smart contract to record data from a user and then relay it back to the user at a predetermined date or in response to an interaction.  Prior to work starting however a development environment will need to be selected and a period spent on set-up and familiarisation.  Whichever development environment is chosen however, the resulting smart contract will be hosted on a DL test network for reasons of cost, performance and security.

			\subsection{Building out to a DApp: Deliverable 2}

			A smart contract without a mechanism to access it falls short of being a DApp -- structurally it could be seen as equivalent to a class within traditional object oriented programming.  The next stage therefore will be to build an infrastructure around that smart contract so it can be accessed by a user via graphical user interface -- e.g. a web browser.  At this point the smart contract can be considered a DApp, although in terms of bare functionality it will have the same fundamentals as previously.

			\subsection{Guidelines: Deliverable 3}

			In the course of building the DApps the literature will be studied to highlight practices which strengthen digital preservation, and reflection will take place on how chosen architectures help or hinder this aim.  Guidelines will then be constructed with the aim of helping future developers consider this when making design choices.

			\subsection{DApp as PIM: Deliverable 4}

			Using the experience gained in building the initial prototype, effort will now switch to designing and building a DApp that includes more functionality and seeks to replicate aspects of a Personal Information Manager.  Clearly the extent of the functionality built will depend on the time available and progress made.  Investigation will also be conducted into relevant standards for such products, e.g. RFC 5545 (iCalendar) \cite{desruisseauxRFC5545Internet2009}, and where interoperability can be achieved with more traditional applications.  Best practices in PIM, which is a field of academic research, as well as a range of applications, will also be considered \cite[p.~51-3]{jonesFuturePersonalInformation2015}.  Noting however; as per Vaishnavi et al's \cite{vaishnaviDesignScienceResearch2019} guidance on design science; novelty (and therefore research contribution) will lay in the design of the artefact, not in its construction.

			\subsection{Evaluation: Deliverable 5}
			As this project is an exploration of unorthodox tools that society might use within digital preservation strategies, the aim is not to develop consumer ready applications but rather to add to the discourse on decentralised alternatives.  It is therefore certain that the deliverables of this project, due to the still evolving nature of DLT and the stability of traditional centralised architectures, will be less performant than the tools currently on offer.

			Resultantly evaluation will not be in the form of user groups or questionnaires, which will inevitably focus on this delta.  Rather the scenarios in Chapter~\ref{chap:Context}, that cover the three pillars of risk (software, hardware and organisation), will be used to evaluate the DApp (Deliverable 4) against the guideline (Deliverable 3).  This will allow an overall judgement on the DApp's effectiveness in securing digital preservation to be provided, and will serve as practical validation for the guidelines. 

\let\textcircled=\pgftextcircled
\chapter{Implementation}\label{chap:Implementation} 

\initial{R}eflecting on the creation of the software deliverables, challenges in doing so and any lessons learnt; is a necessary component of this research.  This will begin with initial choices and development environment set up.  The remaining deliverables: guidelines and evaluation, will be discussed at Chapter \ref{chap:Results}.

\section{Initialisation}
\label{sec:Initialisation}

The hype surrounding DLT, as explored in the introduction, has seen many different DL platforms emerge -- leading to Ethereum's founder, Vitalik Buterin, perhaps not surprisingly, claiming that there is too much competition in the space \cite{fridmanVitalikButerinEthereum2020}.  Clearly the choice of which platform to build on was the initial decision to make: but in doing so it revealed a fundamental dilemma -- is the key criteria one of fitness for development purposes, or for digital preservation.  

These criteria are not the same, and in some ways can even be seen as opposing -- take for example Bitcoin.  By using low level mechanisms to insert small segments of data, and services which patch these together, Matzutt et al \cite{matzuttQuantitativeAnalysisImpact2018} demonstrate that Bitcoin is being used to store arbitrary date other than financial transactions: their 2017 analysis concluding over 1,600 files were being stored, including images.  Non-transactional data can therefore be stored on the Bitcoin blockchain and in theory it could be recorded and later parsed off-chain for a PIM application.  Would data stored on the Bitcoin blockchain be a good choice from a digital preservation perspective? Almost certainly -- as Taleb suggests: ``the longer a technology exists, the longer it can be expected to live'' a phenomenon he calls the Lindy Effect \cite[p.~318]{talebAntifragileThingsThat2012}.  Given that Bitcoin is the longest serving blockchain that has already outlived a number of its peers; analysis by Deloitte suggested that only 8\% of blockchain projects initiated on GitHub were actively maintained \cite{lealtrujilloEvolutionBlockchainTechnology2017}; it seems likely to remain accessible for some time to come.  Even were Bitcoin to suffer a significant drop in value, its data might survive for a considerable period due to the devotion of its adherents and historical significance.  

However the reason for its longevity explains why it would be a poor choice for this project.  It does one thing -- maintain a ledger of transactions -- and aims to do this with a high level of robustness.  Its lack of functionality is therefore less a bug, more a feature -- simplicity leads to less chance of unintended consequences frustrating its main ambition.  This same simplicity, for instance lack of a Turing-complete scripting language, makes it a poor platform to build on.

So clearly digital preservation, although important, could not be the over-riding factor.  The DApp that does not get built is not going to contribute to digital preservation.  From the DL options narrowed down in Section~\ref{sec:ChoosingDL}, a choice was therefore made to build on the Ethereum platform.  At the same time this choice acknowledged that technically there might be superior options. Hedera Hashgraph \cite{bairdHederaPublicHashgraph2019} for instance would almost have certainly proven a more performant choice -- the use of a directed acyclic graph for a ledger means transactions complete quicker resulting in less lag for the end user.  Tezos on the other hand allows formal verification of smart contracts \cite{goodmanTezosSelfamendingCryptoledger2014}, and given that some argue that Ethereum's Solidity language inevitably leads to errors \cite{schransWritingSafeSmart2018}, this might mean a higher quality software product.  Ultimately though as this project's analysis proves Ethereum is far more well established than many of these alternatives: this means it gains both from a digital preservation perspective (i.e. Lindy Effect) and that development on it is achievable: libraries, documentation and active developer communities are available.  

Ethereum has two primary programming languages: Solidity, the original, and Vyper which was introduced to allay security concerns \cite{kaleemVyperSecurityComparison2020}.  Solidity is however a considerably more popular language \cite{moulakhnifVyperGoodAlternative2019} and as a key factor for building on Ethereum was ease of development, Solidity is a logical choice.

The Truffle suite \cite{trufflesuiteSweetToolsSmart2020} of tools was used for local development -- this provided a command line interface that could be used for both deployment and testing using a local blockchain.  Other options, such as OpenZeppelin Software Development Kit \cite{openzeppelinUltimateSmartContract2019}, were considered but Truffle was chosen due to more comprehensive documentation.  The Neovim text editor was used for development due to personal familiarly; the online Remix Integrated Development Environment \cite{ethereumfoundationRemixEthereumIDE2020} was briefly considered but was found unwieldy in comparison.  Given the importance of minimising errors in smart contracts (see Section \ref{sec:Integrity}) the solhint \cite{protofireSolhint2020} and ESlint \cite{openjsfoundationESLintPluggableJavaScript2020} linters were used to validate code during development. 

\section{Smart Contract: Deliverable 1}

The first deliverable stores messages for retrieval at a later date and therefore relies on a schedule to be called in the future -- however there is no way of doing so natively within the Ethereum blockchain.  This is well explained in the initial section of the Solidity documentation:

\begin{quotation}
	``[smart contracts] contain persistent data in state variables, and functions that can modify these variables. Calling a function on a different contract (instance) will perform an EVM function call and thus switch the context such that state variables in the calling contract are inaccessible. A contract and its functions need to be called for anything to happen. There is no `cron' concept in Ethereum to call a function at a particular event automatically.'' \cite{ethereumfoundationContractsSolidityDocumentation2020}
\end{quotation}

This is primarily as Ethereum is designed to run as a ``world computer'' \cite[p.~25]{antonopoulosMasteringEthereum2018}, therefore to introduce such a delay function would involve running a while loop simultaneously on every computer in the network, the transaction fees for which would be very high.  This leaves two options:
			\begin{itemize}
				\item \textbf{Third party}. One solution is to use a third party to monitor Ethereum and detect when a smart contract requires execution.  Alternatives exists for this purpose.  An interesting decentralised initiative is the Ethereum Alarm Clock \cite{chronologicTemporalInnovationBlockchacin2018}.  This is a network of `TimeNodes' where participants are financially incentivised to monitor the Ethereum blockchain for time points where smart contacts should be executed.  If they succeed in performing these executions they are rewarded in cryptocurrency.  Unfortunately at time of writing however, although this system could be used for scheduling Ether transfers, there seemed little emphasis on allowing developers to integrate it with their own smart contracts.  There were alternatives to this however: namely Aion \cite{eth-pantheonAionSchedulingTransactions2018}.  This is a centralised service which could be accessed over Ethereum -- calling an address would result in data being stored locally on a NoSQL database by the provider, which was then used to execute the required function at some point in the future.
				\item \textbf{Client side}.  As discussed at Section \ref{sec:IntroSummary}, this application will follow a three layer architecture, therefore another option is to use the client to query the Ethereum blockchain at regular intervals and receive a response when the correct time has elapsed.  
			\end{itemize}
			The deliverable follows the second of these paths -- the client interrogates the blockchain on a predetermined schedule so ensuring the release of the data at the correct interval.  The resulting smart contract to satisfy the aim was therefore relatively simple and was deployed to the Ropsten test network with two externally accessible functions (the full source code being at Appendix \ref{app:msg-time-store}):
			\begin{itemize}
				\item \lstinline{storeMsg(string memory _storedMsg, uint _unlockTime) public}. The two arguments are the message to be stored, and the time (using unix time) required to be stored until. On calling this function stored messages are entered into a struct which is then pushed into an array mapped against \lstinline{msg.sender} -- i.e. the user's address. This means that Alice will only retrieve her own messages, while Bob will only see messages he has submitted. Eve will see neither.
				\item \lstinline{getMsgTimed() public view returns (StoredData[] memory)}. This returns an array of structs containing all messages which have passed \lstinline{unlockTime}.  Solidity is notable in that although state arrays can be dynamic, memory arrays require a fixed length. Resultantly the returned array will contain empty values for the yet unlocked messages. The client therefore has to check for empty values (specifically \lstinline{id}) prior to presenting to the user, to ameliorate this.
			\end{itemize}
It is of note that although the variables that hold the users’ messages have been specified private, this only relates to whether they can be accessed by other contracts.  Any observer external to the blockchain has visibility of everything inside a contract.  Declaring a variable private simply prevents other contracts accessing that information; but it does not prevent all other external participants seeing that data.

This is a necessity for PIM (see Section \ref{sec:EvalControl}), a fundamental aspect of which is managing privacy \cite[p.~29]{jonesFuturePersonalInformation2012}.  Any real world use of this software (assuming that the user is concerned about privacy) would therefore seek to encrypt information first, or use a DLT which is more privacy minded -- for instance Hyperledger Fabric has privacy of participants data as a core concern \cite{hyperledgerHyperledgerFabricPrivacy2020}.

\section{Message Time Store: Deliverable 2}

With the smart contract established and an understanding gained of data storage and retrieval on Ethereum (Objective 2 at Section \ref{sec:Objectives}); a mechanism is required for the user to interact with it outside of the command line -- the presentation layer as discussed at Section \ref{sec:IntroSummary}.  This featured another decision point -- what client would interact with the DL and how would this be built.  This decision was not however as consequential as that of DL -- the client according to this project's architecture is somewhat transitory -- it is the DL that ensures digital preservation, not the window that the user sees it through.

For interacting with this relatively simple first deliverable a Web browser was chosen as client, due to ubiquity and ease of development speed.  React was chosen to design this front end, this framework being very popular -- named as both the most wanted and loved framework in StackOverflow's 2019 survey  \cite{stackoverflowStackOverflowDeveloper2019}.  Although popularity in itself is not important it had advantages from a research perspective: 

			\begin{itemize}
				\item \textbf{Design science}.  In accordance with the design science principles (see Section \ref{list:designscience}) efforts should be made to communicate research.  Using a well known framework will help others duplicate and build on top of the artefacts.
				\item \textbf{Rapid iterations}.  The more widely used a resource, typically the more development resources, allowing more rapid prototyping.
				\item \textbf{Digital preservation}.  Although the element of the DApp aimed towards longevity is the DL and not the client, the DApp as a whole should be considered a system.  Therefore using components that are well understood and have a large user base will likely ensure a longer life span than more obscure choices.   
			\end{itemize}

Additionally there was also a technical advantage, due to React's ability to reload components without having to refresh the entire web page Document Object Model \cite{facebookReconciliationReact2020}.  This is a useful ability when interfacing a real time client with a blockchain that will only update sporadically depending on block creation.  Although there are several options, \textit{ethers.js} \cite{mooreEthersioEthersJs2020} was used as a JavaScript library to connect the Ethereum DL to React, primarily due to its comprehensive documentation \cite{hayEthereumJavaScriptLibraries2020}.

			\begin{figure}
				\centering
				\includegraphics[width=0.9\columnwidth]{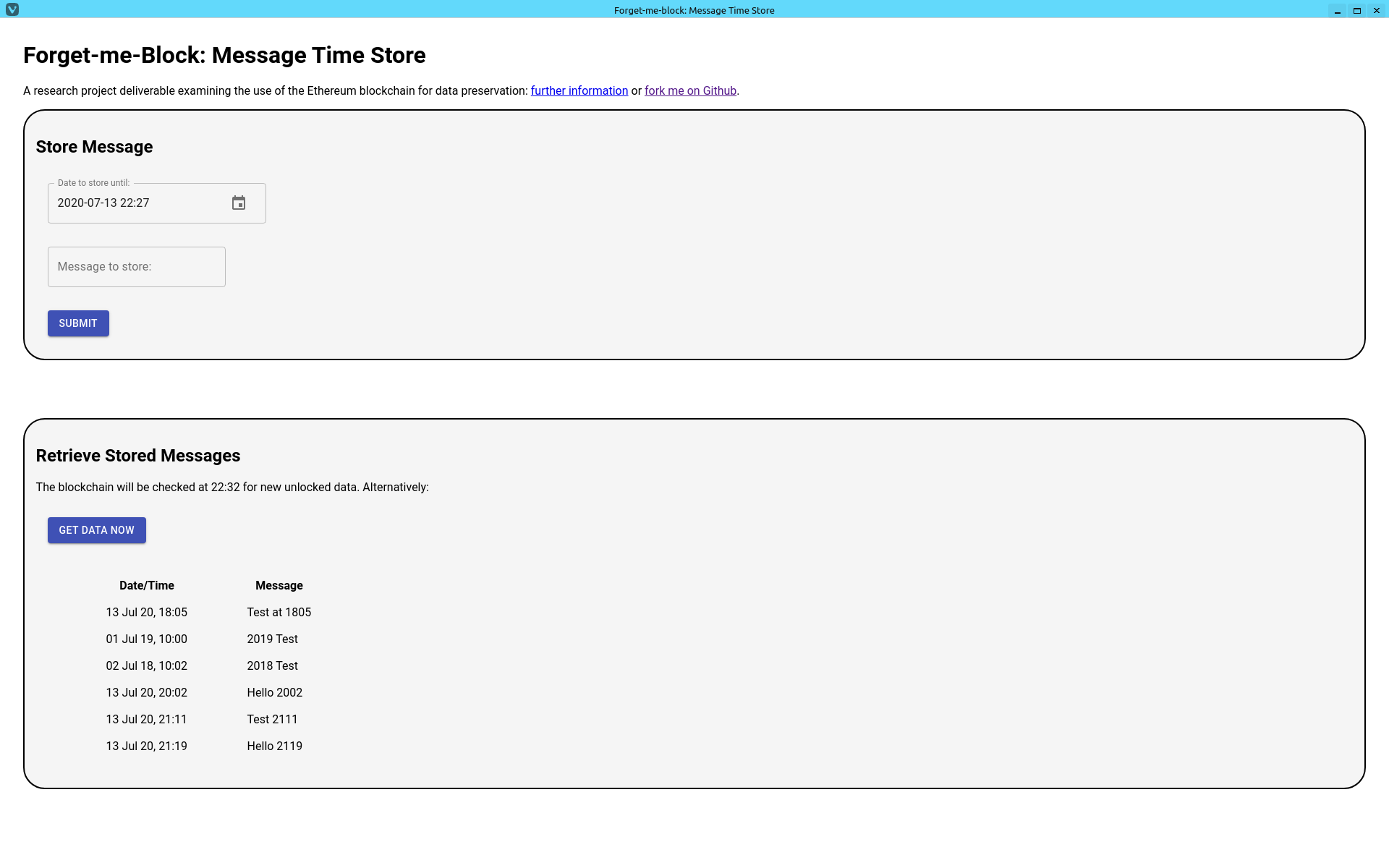}
				\caption[Message Time Store]{Message Time Store (author's own work)}\label{fig:msg-time-store}
			\end{figure}

The resulting client is shown at Figure \ref{fig:msg-time-store} -- this allows a user to store messages on the Ethereum DL with retrieval occurring only after a specified point in time.  Ultimately a DApp had now been developed fulfilling the conditions of Deliverables 1 and 2.  In accordance with the principles of design science both the source code and application itself were made available in combination with a blog post describing its creation \cite{hackmanForgetmeblockMessageTime2020}.

Although this is a relatively trivial application, the process of developing it was a useful learning activity.  Blockchain development, even with a relatively well known language such as Solidity, can be challenging for a number of reasons.  Due to the immaturity of the language, breaking changes are a frequent occurrence (see Section~\ref{sec:PlatformStability}). This meant guidance and advice quickly lost relevance, impeding the learning process.  Breaking changes were not the only challenge -- interfacing with a blockchain is difficult.  Synchronising the user interface with block generation can be problematic, and fault diagnostics suffers as error reporting from the blockchain is not relayed to the developer in an intuitive fashion.  

\section{Ethereum Calendar: Deliverable 4}

Having successfully completed Deliverables 1 and 2 enough experience had been gained to move onto designing software that explored the concerns raised in Chapter \ref{chap:Context} -- that is how might a DApp assist digital preservation in the context of PIM, as per Objective 4 (see Section \ref{sec:Objectives}).

As discussed at Section \ref{sec:pim} the field of PIM is a wide one and covers a variety of applications.  It was decided to focus on a calendar application: partly as this is a key component of the daily working lives of both a clinical psychologist and a marine engineer -- both knowledge workers who require frequent interactions with other professionals and teams.  A calendar also meant that more than one presentation or client layer (see Section \ref{sec:IntroSummary}) could be used to interface with the DL -- namely both a browser and a calendar client (e.g. Mozilla Thunderbird).  This would serve as a better demonstration of the data being held separate from the client layer and in the DL itself.

A considerable number of iterations occurred before ending with a final design.  It is not intended, for reasons of brevity, to cover all these iterations; but rather to examine the final design.  An architectural overview of the system is at Figure \ref{fig:eth-cal-usecase}.

\subsection{CalStore smart contract}\label{sec:CalStore}

\begin{figure}
	\centering
	\includegraphics[width=0.9\columnwidth]{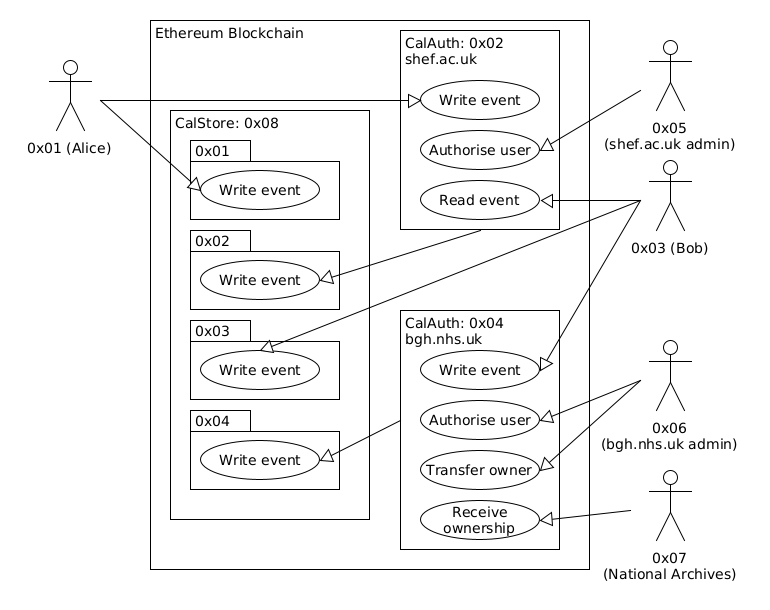}
	\caption[Ethereum Calendar: Use Case Architecture UML diagram]{Ethereum Calendar: Use Case Architecture UML diagram (author's own work)}\label{fig:eth-cal-usecase}
\end{figure}

Figure \ref{fig:eth-cal-usecase} shows that central to the system is the smart contract named \lstinline{CalStore} which is given the example address of \lstinline{0x08}.  The task of this smart contract is to either create, delete or retrieve calendar events for every address that interacts with it.  This builds on the work of the \textit{Message Time Store} deliverable: a caller address mapping is used so each address interacting with the contract is allocated a separate array to store information.  Each new event that an address wishes to store is pushed to this array, while events that are no longer required are deleted and the arrayed shortened.  There are two externally facing (e.g. \lstinline{public})  functions to facilitate creation and deletion: \lstinline{storeEvent} and \lstinline{removeEvent}.  It is of note that there is no requirement to create accounts or users: both the above functions rely on \lstinline{msg.sender}; this being a global variable which indicates which address is calling the contract.  In Figure \ref{fig:eth-cal-usecase} when Alice calls \lstinline{CalStore} then \lstinline{msg.sender} will be equal to \lstinline{0x01}.  The provenance of this (e.g. that Alice actually is Alice) is assured by public-key cryptography.

The remaining two public functions allow an address to retrieve events: \lstinline{getEventsObj} and \lstinline{getEventsIcal}.  An explanation of why there are two functions for the same purpose is required here -- and the advantages and drawbacks of the approach taken.

There is a common standard for calendars: the Internet Calendaring and Scheduling Core Object Specification or iCalendar; which is promulgated by the Internet Engineering Task Force as RFC 5545 \cite{desruisseauxInternetCalendaringScheduling2009}.  The function \lstinline{getEventsIcal} returns text corresponding to a minimal interpretation of RFC 5545 -- given the size of the full RFC and project timeline a complete rendering of the standard was not realistic.  This string, an example of which is at Listing \ref{lst:ical}, is intended to be consumed by calendar clients (e.g. Mozilla Thunderbird).  

The other external function which retrieves events is \lstinline{getEventsObj}, this provides the same information as \lstinline{getEventsIcal}, however in a JavaScript Object Notation (JSON) format.  This function is specifically intended to be used by a web browser calendar application created as part of this project. There is a JSON standard for iCalendar data (jCal -- RFC 7265 \cite{kewischJCalJSONFormat2014}), but this was not implemented.  This was because the browser calendar element relied on a React component \lstinline{react-big-calendar} \cite{quenseReactbigcalendar2020} which had different JSON input requirements.  From the point of view of rapid artefact iteration it was not time efficient to have the Ethereum blockchain produce one JSON format, only for that to be converted to another JSON format in browser.

This design decision however reveals a deeper point as to what logic should be included within a blockchain.  According to the ConsenSys Ethereum Smart Contract Best Practices a developer should: ``only use the blockchain for the parts of your system that require decentralization'' \cite{consensysdiligenceEthereumSmartContract2020}.  Clearly this is not happening here -- the blockchain is being used to deliver the same set of data in two different formats.  If the advice was being followed it would deliver only one version, and the client or presentation layer would be responsible for converting to what was required.

This was considered acceptable for this project: Objective 1 (see Section \ref{sec:Objectives}) was to gain personal experience of implementing DApps, that could only be obtained by writing Solidity code.  Indeed the process of creating appropriate iCal strings in Solidity was a considerable learning process -- for instance converting an unsigned integer or an Ethereum address to a string is not a native capability of Solidity, and therefore functions had to be specifically written to do this.

However a production version of this software might wish to minimise both the calculation that occurs on DL and the results it returns.  In fact both of the jCal and iCal formats could be minimised further.  If we analyse the example iCal from RFC 5545 at Listing \ref{lst:ical} it is apparent there are redundant elements: for instance neither \lstinline{BEGIN:VCALENDAR} and \lstinline{BEGIN:VEVENT} are required if the receiving client is expecting an array of individual events, therefore could be removed from the blockchain function return.

\begin{lstlisting}[language={},caption={RFC 5545 example iCal},captionpos=b,label={lst:ical}]
BEGIN:VCALENDAR
PRODID:-//xyz Corp//NONSGML PDA Calendar Version 1.0//EN
VERSION:2.0
BEGIN:VEVENT
DTSTAMP:19960704T120000Z
UID:uid1@example.com
ORGANIZER:mailto:jsmith@example.com
DTSTART:19960918T143000Z
DTEND:19960920T220000Z
STATUS:CONFIRMED
CATEGORIES:CONFERENCE
SUMMARY:Networld+Interop Conference
DESCRIPTION:Networld+Interop Conference
and Exhibit\nAtlanta World Congress Center\n
Atlanta\, Georgia
END:VEVENT
END:VCALENDAR
\end{lstlisting}

However removing these elements, although potentially aligning with smart contract best practices, could well clash with aims of digital preservation - especially that posed by the `software' threat.  Bodies such as the Research Libraries Group highlight the importance of capturing metadata in the context of digital preservation, as information is only truly understood in context:

\begin{quotation}
... even a simple Web page that contains graphics requires descriptions of the Web environment (browser, etc.), the text (ASCII standard), and the image files. This is a recursive system—all the Representation Information requires additional Representation Information in order to be understood. Representation Information is therefore likely to involve references to other Representation Information elsewhere in the repository and will take the form of a Representation Network. In theory, if the digital object is to remain accessible for the long term, this recursion will stop only when a physical form is encountered, such as a system specification or technical manual \cite{researchlibrariesgroupTrustedDigitalRepositories2002}.
\end{quotation}

Therefore by removing these `redundant' elements we complicate the task of digital preservation; Section \ref{sec:Threat} provides examples of how preservation is frustrated by `software' issues relating to format and metadata.  Instead of the object being directly linked to RFC 5545, we now need further representation information to link the two e.g. in our case part of this representation information would be `append \lstinline{BEGIN:VCALENDAR} to beginning of sequence.'

In this particular conflict between digital preservation and smart contract best practices, it would would seem prudent to lean towards digital preservation -- after all this is the use case that DL is acting in support of.  

\subsection{Hosting}

			\begin{figure}
				\centering
				\includegraphics[width=1\columnwidth]{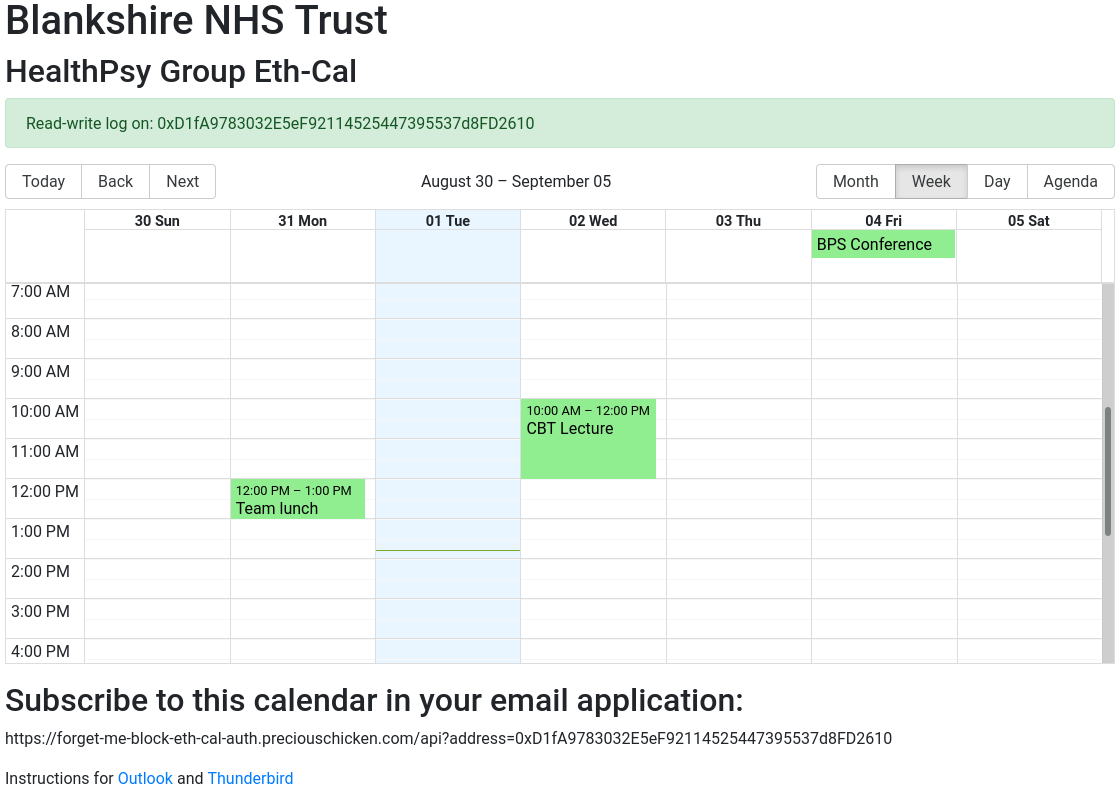}
				\caption[Ethereum Calendar: Web view]{Ethereum Calendar: Web view (author's own work)}
				\label{fig:ethcal-view}
			\end{figure}

			\begin{figure}
				\centering
				\includegraphics[width=1\columnwidth]{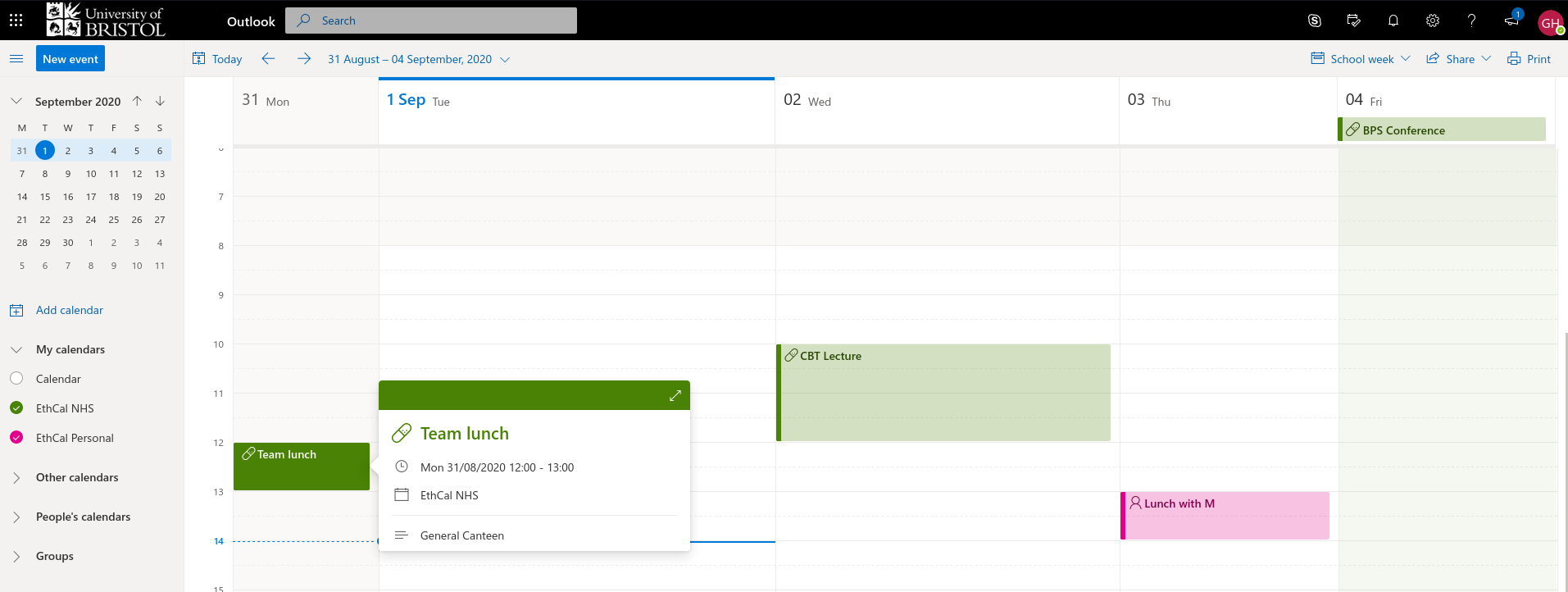}
				\caption[Ethereum Calendar: MS Outlook view]{Ethereum Calendar: MS Outlook view (author's own work)}\label{fig:ethcal-oulook}
			\end{figure}

It is worth dwelling here on the hosting arrangements.  Early iterations of this artefact were hosted on a domestically based web server used for development purposes.  This server hosted two elements: a single page application built using React and a nodejs/express API end point.  The single page application hosted an online calendar, a variant of which is at Figure \ref{fig:ethcal-view}, that users would interact with via their browser -- this was the client that called the function \lstinline{getEventsObj}.  The nodejs/express API end point was an URL that returned a string of text in iCal format via the \lstinline{getEventsIcal} function.  This URL could then be imported into a calendar client, such as Microsoft Outlook, as shown at Figure \ref{fig:ethcal-oulook}.  As one of the aims of design science is research communication, hosting the artefacts on a test-only development server was not sufficient, so the application was migrated to the Vercel cloud platform~\cite{vercelDevelopPreviewShip2020}, necessitating a change of the nodejs/express API into a serverless function. 

One particular aspect of this arrangement involves authentication.  As \lstinline{getEventsObj} is accessed via the browser and Metamask, then \lstinline{msg.sender} can be used to ensure that the caller receives their stored events.  As \lstinline{getEventsIcal} is intended to be accessed via a calendar client this required the user's address (e.g. \lstinline{0x01} for Alice) to be passed as an argument.  Consequently anyone with knowledge of Alice's address could retrieve her calendar.  Ultimately though this was for demonstration only and a solution could be implemented either by setting up a CalDAV server to serve calendars, which requires authentication, or including a Metamask like Ethereum extension into a calendar client which could provide a native authentication process.  This authentication issue also meant that events could only be added via the browser, but again using a CalDAV server would rectify this.  Unfortunately neither solution was possible within the time constraints of this project, although pose an interesting future research direction (see Section~\ref{sec:FutureResearch}).
			
\subsection{CalAuth smart contract}

\begin{figure}
	\centering
	\includegraphics[width=1\columnwidth]{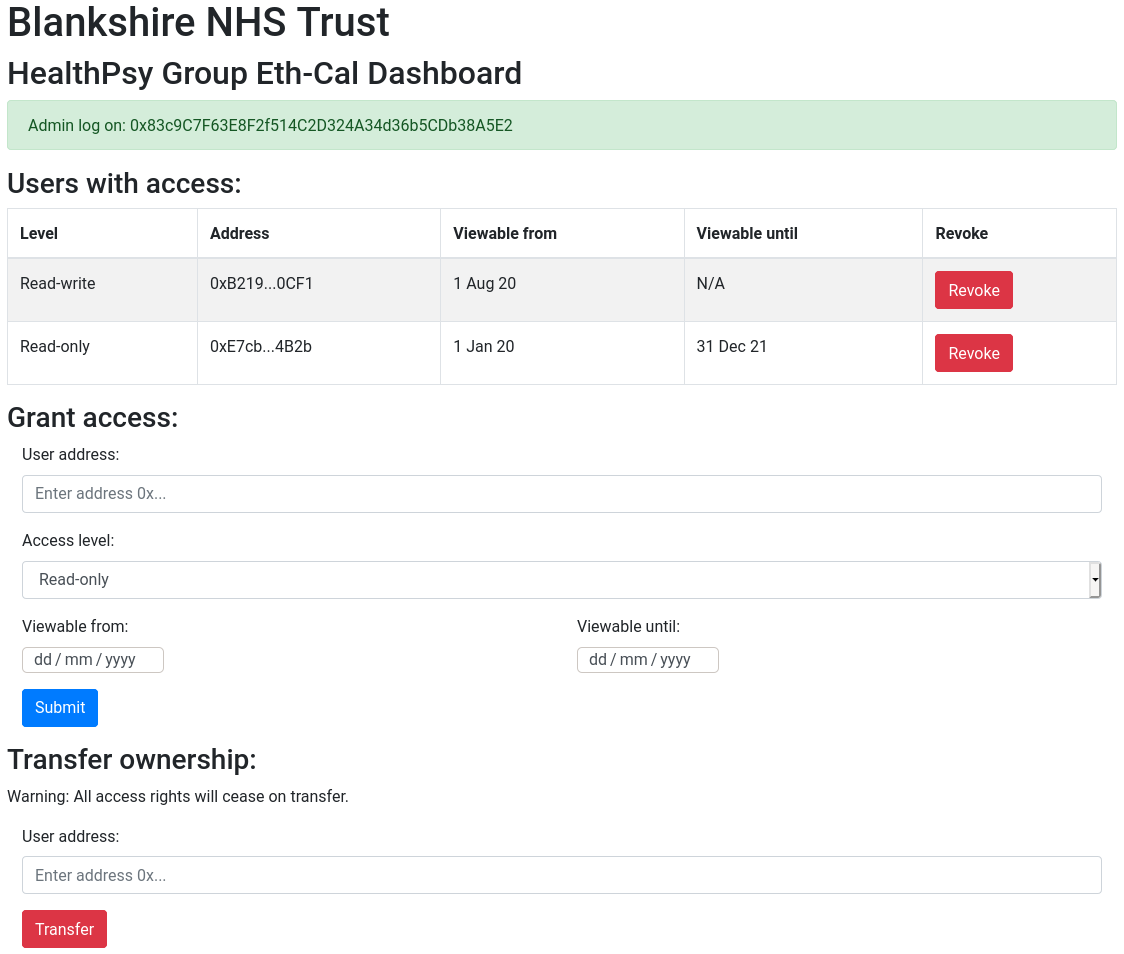}
	\caption[Ethereum Calendar: Administrator's dashboard]{Ethereum Calendar: Administrator's dashboard (author's own work)}\label{fig:eth-cal-admin}
\end{figure}

The second smart contract featured in Figure~\ref{fig:eth-cal-usecase} is \lstinline{CalAuth}.  This is primarily aimed at an organisational level; and relies on Ethereum contracts being able to interact with other contracts in the same way that externally owned accounts (e.g. individuals) are.  \lstinline{CalAuth} is used by an organisation to authenticate users wishing to interact with the organisation's group calendar.  Figure \ref{fig:eth-cal-admin} shows the Authentication screen which is viewable to the owner of the contract -- i.e. the administrator user at the institution who holds the private key corresponding to ownership.  When that administrator user interfaces with the contract they will be presented with a dashboard where they authorise read-only or read-write access to the calendar.  Each of these levels of access can also be date-ranged, i.e. provided with a start and end date.  So referring back to the problem context (see Section \ref{sec:clinpsych}); Bob the clinical psychologist might be granted read-write access to a departmental calendar when he is working at the Blankshire General Hospital.  If his contract is open ended this might have a start date attached to the viewable range, but not an end date.  On leaving the Hospital to join Southampton University the administrator changes his account to read-only with a date range covering his time at the Hospital.

\begin{lstlisting}[language=Solidity,caption={CalAuth transfer ownership function},captionpos=b,label={lst:transfer}]
/// @notice Transfers CalAuth ownership and access management 
/// @dev current owner is revoked ownership and admin
/// @param _newOwner address of new owner
function transferCalAuth(address _newOwner) public onlyOwner {
	grantRole(DEFAULT_ADMIN_ROLE, _newOwner);
	revokeRole(DEFAULT_ADMIN_ROLE, msg.sender);
	transferOwnership(_newOwner);
}
\end{lstlisting}

The contract however not only covers individuals leaving the institution, but also ownership changes to the institution itself.  The \lstinline{CalAuth} contract contains a \lstinline{transferCalAuth} function which transfers both ownership and administrator rights to another party -- but only if the caller of the function is the current owner, as at Listing \ref{lst:transfer}.  Enforcing this requirement is accomplished by the \lstinline{onlyOwner} modifier, part of an OpenZeppelin library.

The \lstinline{CalAuth} smart contract works on the basis that all of \lstinline{CalStore}'s  externally facing functions are duplicated within \lstinline{CalAuth}, taking the same arguments with the same return values.  The presentation layer; be that browser, calendar client or something else; uses the same interface whether it is reading or writing events with \lstinline{CalAuth} or \lstinline{CalStore}, the only difference being the address of the contract it is calling.  The task of \lstinline{CalAuth} is to ensure the user has the correct level of access and then request the appropriate events (depending on date range) from \lstinline{CalStore}.  

\begin{figure}
	\centering
	\includegraphics[width=0.9\columnwidth]{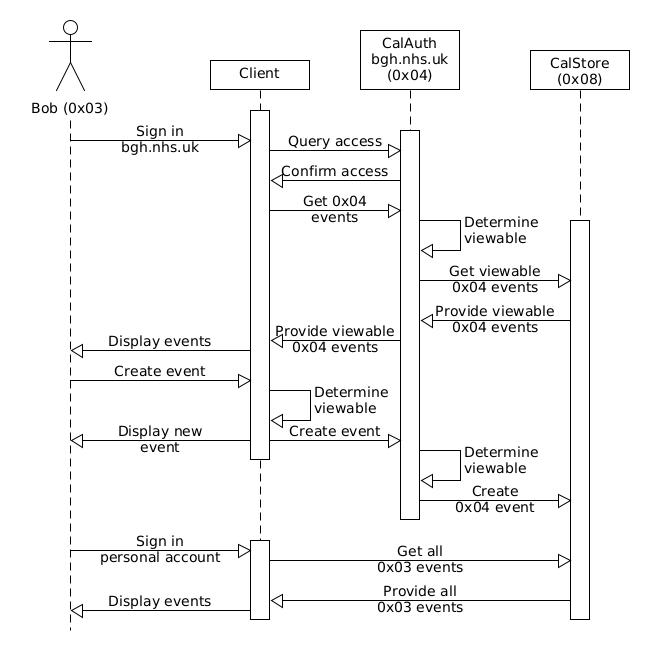}
	\caption[Ethereum Calendar: Sequence UML diagram]{Ethereum Calendar: Sequence UML diagram (author's own work)}\label{fig:eth-cal-sequence}
\end{figure}

Figure \ref{fig:eth-cal-sequence} illustrates this, with Bob using his Ethereum-enabled browser to sign into a web page that displays a Blankshire General Hospital (\textit{bgh.nhs.uk}) departmental calendar to authorised users.  Bob initiates this by confirming his \lstinline{0x03} credentials to \lstinline{CalAuth}, via his client. The smart contract confirms access, relays that to the client and the client in turn requests an array of events from \lstinline{CalAuth}.  \lstinline{CalAuth} determines what events Bob is entitled to view and requests those events from \lstinline{CalStore}, relaying them back to the client.  Of note is that when \lstinline{CalAuth} asks \lstinline{CalStore} for events, it asks for events belonging to the Blankshire General Hospital \lstinline{CalAuth}'s address (\lstinline{0x04}), not Bob's address.  

Should Bob want to create a new event at Blankshire General Hospital a similar process is followed.  As the client is aware of Bob's level of access, having previously received it from \lstinline{CalAuth}, it checks as to whether he can write to the calendar.  If authorised the client requests event creation to \lstinline{CalAuth} and if approved it is subsequently recorded in \lstinline{CalStore}.  Again this is stored at \lstinline{CalStore} using account \lstinline{0x04} belonging to the Hospital, not \lstinline{0x03} belonging to Bob.  A design decision needed to be taken here -- what happens should Bob wish to create an event, but \lstinline{CalAuth} declines; for instance because of a network fault?  Ultimately there were two options: either the client does not display the event until \lstinline{CalAuth} confirms; or the client could display the event immediately, inform \lstinline{CalAuth} that a new event was created and then display an error only on failure.  It was decided to opt for the second of these mechanisms.  Because of synchronisation between client and blockchain there could be a significant delay between event creation and confirmation on the blockchain -- the event not displaying for that timespan would be disconcerting for the user.  Instead it was decided that if the event was refused by the blockchain then the user would receive a warning and the event would remain on the local client (so the user is able to note details) until next browser refresh.

Figure \ref{fig:eth-cal-sequence} also shows Bob interacting with his own personal account once he has finished his session with Blankshire General Hospital.  Here he uses a client, again browser or calendar client, but this time instead of going via \lstinline{CalAuth} his client goes directly to \lstinline{CalStore} which returns all events stored for his address \lstinline{0x03}.  From the perspective of his client it does not matter that he is interacting with \lstinline{CalAuth} or \lstinline{CalStore}, reading and writing events are the same.  Likewise \lstinline{CalStore}'s response is the same whether it is being called by a smart contract or directly by an externally owned account (e.g. an individual).  Figure \ref{fig:ethcal-oulook} illustrates this, the green events are \lstinline{0x04} events from \lstinline{bgh.nhs.uk}'s \lstinline{CalAuth}; the pink event (`Lunch with M') is a personal \lstinline{0x03} event direct from \lstinline{CalStore}.  

Although the example above shows an individual using \lstinline{CalStore} directly and an organisation using \lstinline{CalAuth}, these are not fixed boundaries.  An individual might choose to use a \lstinline{CalAuth} and then give close friends read-only access to their whereabouts, or a couple might use their own \lstinline{CalAuth} to share events; or a carer and the person they support could use one to outline a care plan.  Theoretically \lstinline{CalAuth}'s could also be stacked -- so members of a university might belong to a \lstinline{CalAuth}, and that university's \lstinline{CalAuth} could be pointed to a \lstinline{CalAuth} owned by a federation of universities which then feeds into a \lstinline{CalStore} -- although there would have to be clearly defined advantages to this to outweigh the complexity.

Two variants of client were designed: `open' which featured a web calendar (screenshot not shown) plus API which simulated the user's personal account interacting with \mbox{\lstinline{CalStore};} and `authenticated' which featured a web calendar / admin dashboard plus API which simulated the user interacting with Blankshire General Hospital \lstinline{CalAuth}; both are available online \cite{hackmanForgetmeblockEthereumCalendar2020}.  Both artefacts returned one \lstinline{VCALENDAR} featuring multiple \lstinline{VEVENTS} (see Listing \ref{lst:ical}), although theoretically more sophisticated options are possible.  The next chapter will consider guidelines that have been learnt in constructing these artefacts, and how well they meet the challenges posed by Alice and Bob in Chapter~\ref{chap:Context}.

\let\textcircled=\pgftextcircled
\chapter{Results and Discussion}\label{chap:Results}

\initial{H}aving produced artefacts addressing the problem field discussed; consideration will now be given to establishing guidelines, based on both reflection and the literature, aimed at developers and researchers designing DApps to address digital preservation.  Explanatory comments on the guidelines will be provided, before evaluating the DApp against the guidelines in the context of the scenarios at Chapter \ref{chap:Context}. 

\section{Working towards guidelines}

\begin{figure}
				\centering
				\includegraphics[width=0.6\columnwidth]{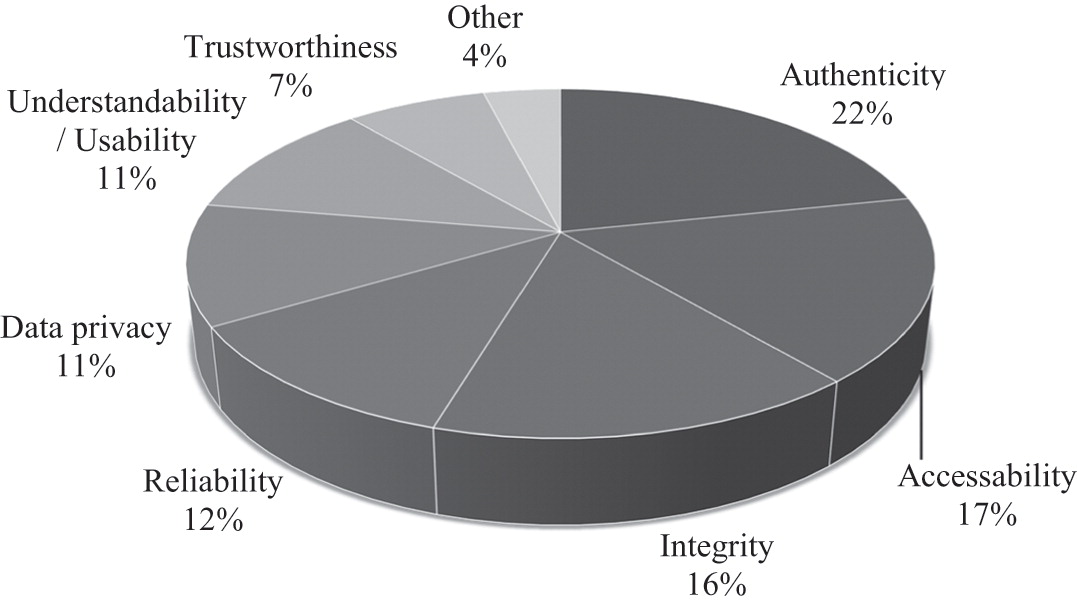}
				\caption[Burda \& Teuteberg's attributes relevant for digital preservation]{Burda \& Teuteberg's attributes relevant for digital preservation \cite{burdaSustainingAccessibilityInformation2013}}\label{fig:burda_dp}
			\end{figure}

Consideration was first given to an examination of the literature on digital preservation in constructing the guidelines, so to understand common themes of threat and mitigation (as per Objective 7 at Section \ref{sec:Objectives}).  A key source here was Burda \& Teuteberg's systematic literature review \cite{burdaSustainingAccessibilityInformation2013}, which surveyed 122 publications for themes in digital preservation.  Figure \ref{fig:burda_dp} shows the results of their survey as to how often non-functional requirements (i.e. attributes) were raised.

It is worth defining these terms.  The most mentioned terms were `authenticity' (an object is what it claims to be), `accessibility' (whether on object can be retrieved in a timely and continuous manner) and `integrity' (whether data has been modified, either accidentally or maliciously).  Following that the next most popular terms were `reliability' (digital objects survivability through a system's lifetime), `privacy' (preventing unauthorised access), `usability' (ability to use the digital object in the manner intended) and `trustworthiness' (ability to demonstrate digital preservation has been conducted competently).  After that in the `other' category came performance, scalability and cost-effectiveness.  Clearly these items at time overlap -- `accessibility' and `reliability' are not mutually exclusive: if a digital object no longer survives then it is de facto non-accessible. These terms, based on this comprehensive study, will therefore serve as the bedrock of our guidelines.

\begin{figure}
	\includegraphics[width=1.0\columnwidth]{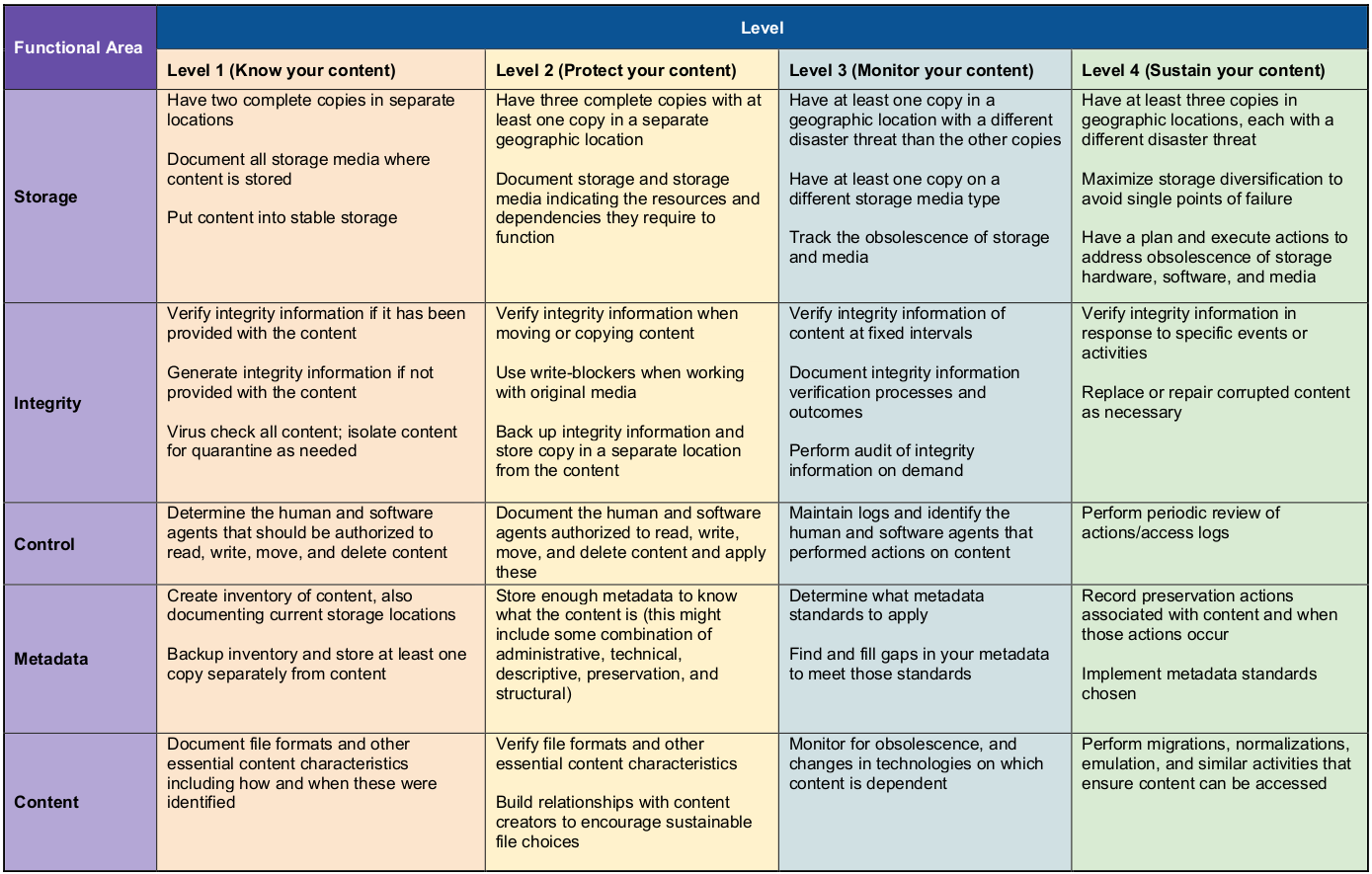}
	\caption[NDSA 2019 Levels of Digital Preservation Matrix]{NDSA 2019 Levels of Digital Preservation Matrix \cite{kussman2019LevelsDigital2020}}\label{fig:nsda_levels}
	\centering
\end{figure}

As well as reviewing the literature, it is also useful to survey advice from those responsible for the practicalities of digital preservation.  One such body is the US National Digital Stewardship Alliance (NDSA), a consortium of institutions committed to digital preservation.  Figure \ref{fig:nsda_levels} shows their Levels of Digital Preservation matrix which is meant to act as a guide to organisations attempting to actively pursue a strategy of digital preservation.  Clearly there is considerable cross-over with the attributes that Burda \& Teuteberg have listed: some concepts directly map, i.e. `integrity', while others are synonyms, i.e. `data privacy' for `control'.  As might be expected there is more of an emphasis on the practicalities of digital preservation than in Burda \& Teuteberg: for instance maintaining an inventory of what is stored and the specifics of file formats.

\begin{figure}
	\centering
	\includegraphics[width=1.0\columnwidth]{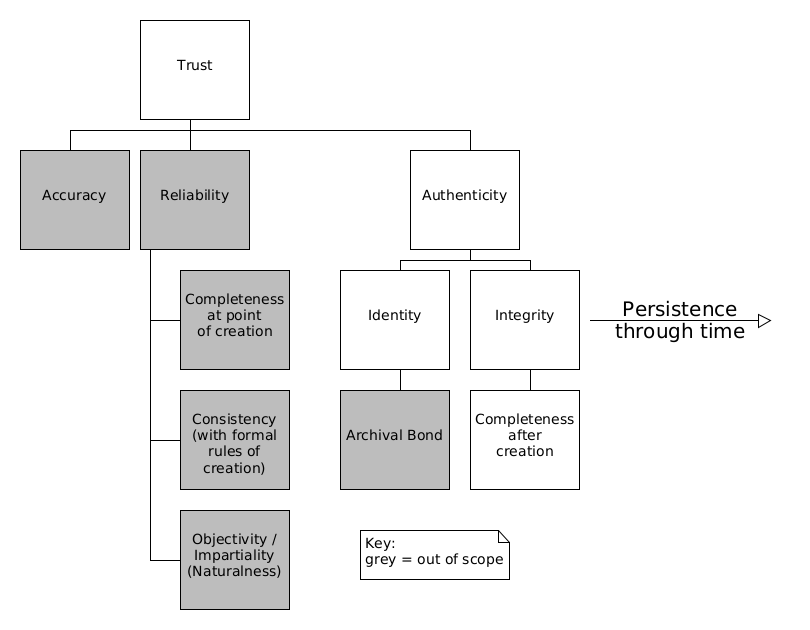}
	\caption[Lemieux's blockchain evaluation using the Archival Theoretic Recordkeeping Framework]{Lemieux's blockchain evaluation using the Archival Theoretic Recordkeeping Framework \cite{lemieuxBlockchainDistributedLedgers2017}}\label{fig:lemiuex}
\end{figure}

Any discussion of guidelines cannot be complete without examining the work of Lemieux \cite{lemieuxBlockchainDistributedLedgers2017}, who evaluates the use of DLT via the Archival Theoretic Recordkeeping Framework as at Figure \ref{fig:lemiuex}.  Due to this framework's roots in archival science, not digital preservation, the concepts do not always map across to DLT, as illustrated by the greyed out concepts in Figure \ref{fig:lemiuex}.  It is worth examining the divergence: `accuracy' relates to truthfulness or correctness, which of course a DL in itself cannot establish.  While `reliability' in the digital preservation context relates to survivability in an information system, within archival science it relates to `completeness;' to use Lemieux's example a sale of land contract missing a signature would be considered incomplete and therefore unreliable.  `Archival bond' relates to the record's provenance; ultimately how did the item end up as a record.  It is worth noting that Lemieux's decisions as to this mapping rely on the proviso that the DL would be used to store a hash of the digital object and not the digital object itself; if this is not the case, as Lemieux admits, her mapping might differ.  The terms `identity' and `integrity' which combine to form `authenticity' and from then `trust' are similar to the previous sources and do not require further explanation.  

\section{Guidelines: Deliverable 3}

Following consideration of these sources, combined with reflection on building the artefacts at Chapter \ref{chap:Implementation}, guidelines aimed at other developers and researchers designing DApps to address digital preservation are at Table \ref{table:guideline_criteria} (so contributing towards Objectives 5 and 6 at Section \ref{sec:Objectives}).

The guidelines invite the designer to score their DApp against five criteria; \textit{storage}, \textit{accessibility}, \textit{integrity}, \textit{usability} and \textit{control \& identity}; with the DApp being awarded a score for each criteria from 1 (worst) to 5 (best).  Each score is associated with a boolean statement that evaluates to true or false.  The user should answer questions from left to right and the score awarded should be the last answered `true' for that criteria.  So for example if in \textit{accessibility} a DL has an active developer base (1) and an open source ecosystem (2), but does not have an active research community (3), then the total score for accessibility should be 2, regardless of answers for scores 4 and 5.

The maximum score therefore is 25 points.  It is the view of this research that the more points a DApp scores the more likely it is to meet the challenges of digital preservation.

Due to DLT's ``conceptual fuzziness'' \cite{osternBlockchainResearchDiscipline2019} and the myriad challenges set out at Chapter \ref{chap:Context}, this is never going to be an exact science however and some flexibility may be required.  For instance under \textit{integrity} it might be the case that all other questions are true, apart from the `industry standard libraries' as there are none available for that particular use case.  Here discretion may be used and a score higher than 2 awarded.  Caution is necessary however -- if the use case for a DApp is so new that there are no applicable libraries that in itself is a warning sign.  
	\newgeometry{bottom=5mm} 
\begin{landscape}
	\begin{table}
		\centering
		\renewcommand{\arraystretch}{2}
		\begin{tabular}{ m{5em} m{11em} m{11em} m{11em} m{11em} m{11em}  } 
			\toprule
			Criteria & 1 (Worst) & 2 & 3 & 4 & 5 (Best) \\
			\midrule
			Storage 
			& Solution has sufficient capacity for maximal use case. 
			& Data hosted in more than one location, in standardised format.
			& Data hosted in multiple global locations. 
			& Hosted on permissionless DL / DDb < 10,000 nodes.
			& Hosted on permissionless DL / DDb > 10,000 nodes. \\
			Accessibility 
			& DL has active developer base.
			& DL has open source ecosystem.
			& DL has active research community.
			& DL actively employed for real world use cases, including corporates.
			& DL actively employed for real world use cases, including government. \\
			Integrity
			& Source code test coverage > 80\%. 
			& DApp uses industry standard libraries.
			& DL sufficiently mature and stable.
			& Smart contracts have been audited.
			& Smart contracts are formally verified.\\
			Control \& Identity	
			& DApp identifies users. 
			& DApp encrypts personal data. 
			& DApp permits transfer of ownership. 
			& DApp implements RBAC. 
			& DL / DApp implements user-friendly self-sovereignty \\
			Usability	
			& User can complete use case.
			& DApp is performant. 
			& DApp scales for multiple users. 
			& DApp uses human readable attributes.
			& DApp uses native interface.\\
			\bottomrule
		\end{tabular}
		\caption[Digital preservation guidelines for DApps]{Digital preservation guidelines for DApps (author's own work)}\label{table:guideline_criteria}
	\end{table}
\end{landscape}
	\restoregeometry
Likewise some scores are subjective -- within \textit{accessibility} what exactly constitutes an `active research community' -- the production of \textit{x} papers per year?  Some value judgements are therefore required.  This also illustrates that although some scores fall within the DApp developer's influence (e.g. using industry standard libraries) a considerable number do not -- any one development team is unlikely to change how active a research community is.  The aim here is to give an assessment of how likely a DApp is to succeed in digital preservation over the long term, not simply to certify a best effort endeavour.

Ultimately the guidelines are intended as a hand rail and not a straight jacket to all those interested in the field; it is hoped they will genuinely contribute to a practical understanding of how DLT might best serve digital preservation.

Each criteria will now be covered in turn, with each beginning with a general discussion and then providing comment, where required, on the guidelines.  Following this the Ethereum Calendar DApp will be evaluated against these guidelines at Section \ref{sec:DiscEval}.

\subsection{Storage}\label{sec:storage}

The initial step in digital preservation is the system's ability to store information prior to anything else.  The ability to store data on the Ethereum DL itself is extremely limited: an Ethereum smart contract can only store 24KB \cite{buterinEthereumImprovementProposal2019}.   There are alternative means to introduce more storage on the Ethereum blockchain, e.g. the Diamond standard \cite{mudgeEthereumMaximumContract2020} or a stateless smart contract \cite{childs-maidmentStatelessSmartContracts2017}.  However as Section \ref{sec:IntroSummary} discusses if the objective is to store data other than transactions it is optimal to do so on a third layer decentralised database (DDb) e.g. IPFS \cite{benetIPFSConentAddressed2014}, BigchainDB \cite{bigchaindbBigchainDBBlockchainDatabase2018} or Storj \cite{storjlabsStorjDecentralizedCloud2018}.  The literature already covers a number of design science examples where these platforms have been used in conjunction with DLs such as Ethereum \cite{zichichiEfficiencyDecentralizedFile2020, dieboldSelfSovereignIdentityUsing2017, wangBlockchainBasedFrameworkData2018, zhangDecentralizedModelSpatial2020}.  DDbs are however in the same state of infancy as DLs, potentially more so; for instance a comprehensive mapping of IPFS found that the majority of its nodes were unreachable \cite{henningsenMappingInterplanetaryFilesystem2020}.  Any DApp developer therefore needs to be mindful of their ability to store data within their chosen architecture.

This criteria also refers to how the data is stored -- where possible data should be formatted in accordance with a standard (e.g. RFC 5545 in the case of this DApp).  Although this does not guarantee that future communities will be able to read the information stored, a format that is more commonly recognised will increase the likelihood of successful retrieval.

Scoring for this criteria relates back to the architectural discussion in Section~\ref{sec:IntroSummary} and considers how storage for DApps can be centralised or decentralised (Figure \ref{fig:bigchaindb}) and if the latter as permissioned or permissionless DL or DDbs (Figure \ref{fig:walport}).
			
\begin{enumerate}
	\item \textbf{Solution has sufficient capacity for maximal use case}.  Given the lack of storage capacity within the Ethereum DL, and others, this is the most basic requirement, regardless of how or where data is stored. 
	\item \textbf{Data hosted in more than one location, in standardised format.}  Storing data in more than one location is an obvious first principle of digital preservation.  Data should also, where possible, be stored in accordance with a relevant standard (e.g. RFC 5545).
	\item \textbf{Data hosted in multiple global locations}.  This score could also apply to those who choose to use a centralised Db or permissioned DL.
	\item \textbf{Hosted on permissionless DL / DDb < 10,000 nodes}.  Although estimating the footprint of any given DLT is inexact, a 2018 study \cite{kimMeasuringEthereumNetwork2018} placed Bitcoin with 10,454 nodes and Ethereum with 14,545; while a 2020 study \cite{henningsenMappingInterplanetaryFilesystem2020} found 3,253 reachable nodes on IPFS.  Ten thousand nodes is therefore proposed as a relatively arbitrary dividing line between a major and minor DL / DDb -- noting that this will likely change over time.  The assumption being the larger the distributed network the more it is likely to further digital preservation.    
	\item \textbf{Hosted on permissionless DL / DDb > 10,000 nodes}.  It is the view of this research that this option does most to preserve digital preservation -- decentralisation mitigating the `organisational' risk, while a wide distribution the `hardware' risk.
\end{enumerate}

\subsection{Accessibility}\label{sec:accessibility}

The larger challenge, though is not whether data can be stored but whether it can be accessed over the long term, which relies on the success of the underlying DL.  

It has been argued that Nakamoto's genius lay not in invention, but rather combining already existing components \cite{narayananBitcoinAcademicPedigree2017} which intertwined economics and computer science.  The proof-of-work consensus mechanism incentivises participants to follow the rules of the system and in doing so strengthen it.  Indeed cheating the system would require such massive resource allocation that it is unlikely one would recoup the costs before the currency's value would disappear on discovery of a successful exploit \cite{7163021}.  Ultimately though the use case for Bitcoin is simple -- allowing individuals to own ``electronic cash'' \cite{nakamotoBitcoinPeertoPeerElectronic2008} not intermediated by either banks or governments.  Whether or not one agrees with that, it is a clear aim.

However as we move away from Bitcoin -- which is very limited outwith electronic cash -- an understanding of the value proposition becomes less certain.  As already established Ethereum is by far the most popular `Turing complete' DL, portraying itself as a ``world computer'' \cite[p.~25]{antonopoulosMasteringEthereum2018} -- we might therefore expect there to be an explosion of innovation on such a platform.  However if we look at Oliva et al's recent study of Ethereum \cite{olivaExploratoryStudySmart2020} we see that a tiny fraction (0.05\%) of smart contracts are currently the target of 80\% of transactions on the system.  Of those highly popular smart contracts five out of the top ten belong to currency exchanges and a high proportion of the remnant concern token activities such as Initial Coin Offerings.  A proportion of that will be plain speculation -- the apotheosis of which may well be the ``Useless Ethereum Token.''  This advertised itself with the statement ``you're going to give some random person on the internet money, and they're going to take it and go buy stuff with it'' \cite{anonUselessEthereumToken2017}  and yet still managed to raise US\$300,000 \cite{markUselessEthereumToken2018}.  This imbalance towards highly speculative tokens may not be a healthy ecosystem for a platform described as a ``world computer.''

A decentralised platform has to have some method of intrinsic value creation -- in other words it needs to be able to incentivise people to run the software.  If the majority of activity on the Ethereum DL is the buying and selling of tokens -- will that continue to be a sustainable driver of incentivisation over the long term?  Incentivisation is also itself a finely balanced act.  The higher the fees the more attractive it is for providers to add resource to a system (i.e. compute, storage), but those higher fees will discourage users of those resources.  Nor is it a given that the laws of demand and supply will result in an equilibrium point where such a system is viable (e.g. the price the consumer is willing to bear exceeds the fixed and variable costs of suppliers).  Coupled with this uncertainty is the reality that much of the competition, especially in the field of PIM, is already `free' at the point of use.   Trends however change: the rise of privacy focussed services such as Protonmail \cite{infosecuritymagazineUberSecureProtonMailBeta2014} could suggest an increasing consumer realisation that `free' means they are ``the product, not the customer'' \cite{solonYouAreFacebook2011}.  Additionally it might be the case that the `fee' consumers pay to use these systems is by sharing their own under-utilised resources (e.g. hard drive space) \cite{protocollabsFilecoinDecentralizedStorage2017}, which may be more palatable.  

The costs associated with deploying and using the smart contracts in this research are shown at Table \ref{table:fees} -- Sterling comparisons have been made noting that Ether on the Ropsten test net has no monetary value.  An important point in favour of digital preservation on the Ethereum DL is that  costs are imposed for `writes' and not `reads,' this model has been duplicated by other third layer decentralised storage proposals \cite{lambertSAFENetworkNew2014}.  Unlike a traditional centralised provider where typically you pay for access and when you stop paying you lose access, with much of the decentralised web you pay once only.  This strengthens digital preservation as if the entity who uploads loses the capacity to pay (e.g. they disestablish), or even if they merely lose interest in the project, the data remains on the DL for others to access.

\begin{table}
	\centering
	\begin{tabular}{ l r r } 
		\toprule  
		\multicolumn{1}{c}{Function} & 
		\multicolumn{1}{c}{Ether} & 
		\multicolumn{1}{c}{Sterling}  \\ 
		\midrule 
		Deploy contracts  & $\Xi$ 0.11966 & \pounds 41.68 \\ 
		Create event & $\Xi$ 0.00021 & \pounds 0.07 \\ 
		Delete event & $\Xi$ 0.00009 & \pounds 0.03 \\ 
		Transfer ownership & $\Xi$ 0.00040 & \pounds 0.14 \\ 
		\bottomrule
	\end{tabular}
	\caption[Ethereum Calendar transaction costs on Ropsten network]{Ethereum Calendar transaction costs on Ropsten network on 1 September 2020 (author's own work)}
	\label{table:fees}
\end{table}

Permissioned DLs (see Section \ref{sec:Overview}) tend to be free of speculation and do not require an internal financial incentive due to their governance process -- i.e. normally being established by a consortium.  Although this has some advantages, it reduces the key digital preservation advantages of a DL by tying the longevity of the data to the fortunes of the entities that provide the permissions.  It can be argued though that some digital preservation advantages are maintained as the fortune of the DL tends to depend on a consortium (which theoretically can welcome new members), rather than just one centralised entity.

Both permissioned and permissionless DL however suffer from the accusation that they, or the DApps built upon them, are often simply `vapourware' -- initiatives that are set up for reasons of publicity rather than solid business reasons \cite{chengRidingBlockchainMania2019}.  Several studies support this notion: MERL Tech examined 43 DLT international development use-cases finding little evidence of concrete results \cite{burgBlockchainInternationalDevelopment2018}.  Similarly Naqvi and Hussain \cite{naqviEvidenceBasedBlockchainFindings2020} analysed 517 DLT projects and found less than 2\% demonstrated substantial evidence.  Although this latter study appears comprehensive, it is noticeable that the lead author is the Founding Editor-In-Chief of the journal publishing it: the journal does have a statement of policy on submissions by members of the editorial team, though the association is not mentioned in the article itself \cite{grafBestPracticeGuidelines2006}.

A key question for any developer therefore considering digital preservation is what is the long term value proposition of the DL platform they are building on.  Given the novelty of this field, this appears too early to tell in many cases.

This project's scoring for accessibility will therefore consider the wider software ecosystem surrounding a DL.  The logic being that a healthy ecosystem will likely point to a mechanism of intrinsic value creation that is working in a sustainable manner.  This approach however does accept that some subjectivity is unavoidable.

\begin{enumerate}
	\item \textbf{DL has active developer base}.  To be gauged via Git repository commits, question and answer forums (e.g. StackOverflow), online tutorials, etc.  Ultimately participation between a number of different entities lies at the heart of a healthy ecosystem \cite{denhartighMeasuringHealthBusiness2013}.
	\item \textbf{DL has open source ecosystem}.  Although closed source projects can be equally valid, open-source projects have an advantage in exposing vulnerabilities to a greater number of eyes, and particularly important for digital preservation is that development does not rely solely on the goodwill of the original maintainers \cite{watsonBusinessOpenSource2008}. 
	\item \textbf{DL has active research community}.  Scrutiny from peer-reviewed journals is likely to reduce errors and promote best practice.
	\item \textbf{DL actively employed for real world use cases, including corporates}.  Addressing the concerns above, developers should look to build on DLs that have already established themselves by solving real world problems rather than producing `vapourware' or DApps limited to test cases.  The involvement of larger corporates using DL for actual business processes (rather than pilots) is a key indicator of this.
	\item \textbf{DL actively employed for real world use cases, including government}.  As governments typical adopt novel technologies more cautiously than the private sector, this is a further endorsement that a DL is a long term option.  Again this should be actual established use, rather than pilot studies.
\end{enumerate}

\subsection{Integrity}
\label{sec:Integrity}

Next to being able to store and access data, a developer must be mindful that accident or malign intent does not change the data they have stored.  Developing within the DL ecosystem brings a number of new challenges.

Nikolic et al \cite{nikolicFindingGreedyProdigal2018} sampled 3,759 Ethereum smart contracts using a validation tool and found 89\% had some form of exploitable vulnerability.  This is not an isolated example: Luu et al \cite{luuMakingSmartContracts2016} found in their analysis of 19,336 Ethereum smart contracts that 46\% displayed vulnerabilities -- a lower figure but still significant.  Unfortunately neither study performs similar analysis against other programming languages, so it is difficult to categorically state this is specific to Solidity.  However Oliva et al \cite{olivaExploratoryStudySmart2020} have found that Solidity smart contracts display more ``complexity characteristics'' than programs in other popular languages.  Porru et al \cite{porruBlockchainorientedSoftwareEngineering2017} have taken this further and stated that much of blockchain development is ``unruled and hurried'' with consequences for software quality.   Taken together this has led to calls for an increased engineering mindset in DL development \cite{destefanisSmartContractsVulnerabilities2018, porruBlockchainorientedSoftwareEngineering2017} as well as proposals of new tools \cite{tikhomirovSmartCheckStaticAnalysis2018, luuMakingSmartContracts2016} and even new languages to improve integrity \cite{schransWritingSafeSmart2018}.

This project does not have the scope to present a review of this material, however reflection on creating this DApp combined with a study of the literature raises several key areas that designers should be aware of:

\subsubsection{Use of standards} 

Given the propensity for errors when developing with Solidity, and by extension other DL programming languages, developers should seek to use standardised libraries provided by reputable organisations.  As Kondo \cite{article} notes there is an expectation that developers should use code from reputable sources, giving OpenZeppelin as one such example which is ``devoted to creating secure libraries.''  

A key recommendation therefore is for developers to look for DLs and languages that already have an existing collection of reputable libraries, and then use them in their DApps.

\subsubsection{Platform stability}\label{sec:PlatformStability}

Reflecting on the experience of building a Solidity DApp, the pace of the language's development was keenly felt, especially regards breaking changes.  A particularly egregious example was the change in the format of the constructor function, introduced in release 0.5.0~\cite{ethereumfoundationSolidityV0Breaking2018}; the fact that such fundamental aspects of the language are still being debated illustrates the degree of change.  As Tikhomirov et al \cite{tikhomirovSmartCheckStaticAnalysis2018} critique: ``advice on secure Ethereum programming practices is spread out across blogs, papers, and tutorials... [much of which is] ... outdated due to a rapid pace of development in this field.''

This is not just a problem for code which has not yet been compiled, it is also an issue for running instances of smart contracts.  A lead Ethereum developer, Wei Tang, has suggested that the community does not give enough emphasis to backwards compatibility \cite{janusEthereumHardFork2019} and has proposed changes which would promote this \cite{tangVersionlessEthereumVirtual2019}.  

Building on this one might also consider the general `culture' of the platform towards stability.  A prime example of this is the notorious DAO attack that took place on the Ethereum DL \cite{dupontExperimentsAlgorithmicGovernance2017}.  Following this those with power over the direction of the platform decided that a hard fork, which compensated victims, was preferable to a strict adherence to a `code is law' philosophy \cite{filippiBlockchainTechnologyRegulatory2016}.  This contrasts sharply with the Bitcoin DL where the core developers see continuity as a guiding principle.  Regardless of the merits of either approach, from a digital preservation perspective a platform with a culture of stability is to be preferred.

As an aside there is a case to argue that a language that is preserved in aspic, such as COBOL, might lead to a lack of interest and therefore reduce the popularity (and hence longevity) of the platform.  This should not be overemphasised however -- the C programming language is incredibly stable \cite{iso-9899.infoStandard2019}, and at the same time is a mainstay of projects such as Linux (the most popular server operating system worldwide \cite{finleyLinuxTookWeb2016}).  

A developer concerned about data prevention therefore has to be mindful as to the stability of the platform they are building on.  A frequently changing landscape will make it more difficult to build stable artefacts, and a lack of backwards compatibility does not bode well for digital preservation.

\subsubsection{Designing for forever} 

The Ethereum platform has gained some notoriety for the ease with which mistakes are made \cite{dupontExperimentsAlgorithmicGovernance2017}, reflection on this project's deliverables bore this out.  At one point during development is was possible for a logged on user to see another user's events (e.g. Alice would see Bob's events appearing on her calendar).  The error here was a state variable was appended to by each user and pushed into each accounts array, however the variable was not being correctly cleared between logins.  This was easily corrected -- however it illustrates a mindset change required.  Frequently when deploying software we expect the execution to be instantiated either for a limited time (as long as a user is logged on) or with many different instances (one per user for instance).  Developing a smart contract is the exact opposite -- we have one instance that is designed to run forever and that instance is externally accessible to anyone with access to the internet.  

Although it is now possible to make changes to a deployed contract \cite{openzeppelinUpgradingSmartContracts2020}, the default position is that one does not  \cite{wangBlockchainEnabledSmartContracts2019}.  And even if theoretically possible it is often operationally very difficult, as the developers of a popular Ethereum game reveal what a bug fix would mean in practice:

\begin{quotation} 
We could fix Unexpected Kitty Fleas by making a change to a single line of solidity code. However, if we push the fix through, all Siring Auctions would need to be cancelled and reposted (that’s over 40,000 auctions), and our users wouldn’t be able to sire their cats for 12+ hours.

Simply put: it makes more sense for us to issue refunds at a loss AND donate the entirety of the overpayment to charity than it does to fix such a minor issue (and disrupt a key part of the game for nearly a full day) \cite{cryptokittiesUnexpectedKittyFleas2018}.
\end{quotation}

This idea therefore that the instance is forever, although potentially useful for data preservation, does provide integrity challenges when developing systems.

The challenges above lead to a number of best practices appearing in the guidelines:

\begin{enumerate}
	\item \textbf{Source code test coverage > 80\%}.  Test suites are available within Ethereum \cite{reaCodeCoverageSolidity2016} and other DLs, and adopting a testing mindset is likely to increase integrity.  This also includes thorough code documentation \cite{parnasSoftwareAging1994}.
	\item \textbf{DApp uses industry standard libraries}.  As above.
	\item \textbf{DL sufficiently mature and stable}.  A DL where the language or platform is undergoing frequent breaking changes is unlikely to lead to DApp integrity; although admittedly determining maturity and stability is a judgement call.
	\item \textbf{Smart contracts have been audited}.  Several external parties; e.g. OpenZeppelin, ConsenSys; exist to independently audit DApps. 
	\item \textbf{Smart contracts are formally verified}.  Although not all programming languages support this, formal verification is likely to reduce error rates. \cite{rosuFormalDesignImplementation2018}. 
\end{enumerate}

\subsection{Control \& Identity}

The guidelines combine control and identity as these two facets are closely related.  Who one gives access to, is closely related to being able to correctly identify users.

The Ethereum Calendar deliverable implemented a mechanism, via \lstinline{CalAuth}, where only authorised addresses could write to the system.  Although there was a mechanism to control read-access this was demonstration only -- the default position of Ethereum is all data is readable by external parties, unless it has been encrypted.  This can be addressed however: Wang et al \cite{wangBlockchainBasedFrameworkData2018} demonstrate sharing distributed encrypted data through the use of Ethereum and IPFS working in concert.  Likewise, as mentioned previously, some DLs, e.g. Hyperledger \cite{hyperledgerHyperledgerFabricPrivacy2020}, have privacy as a fundamental concern.  

\begin{figure}
	\centering
	\includegraphics[width=0.9\columnwidth]{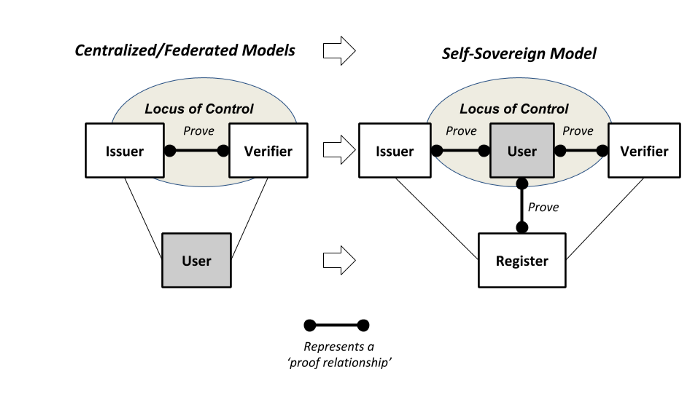}
	\caption[Bouma's self-sovereign model]{Bouma's self-sovereign model \cite{boumaSelfSovereignIdentityShifting2019}}\label{fig:bouma}
\end{figure}

A more complex issue though as it relates to control and identity, as discussed in Section \ref{sec:Overview}, is DLT's distrust of centralised entities or as Szabo would refer to them: ``security holes'' \cite{szaboTrustedThirdParties2001}.  The direction of travel in DLT is to replace third parties with the `self-sovereign individual' \cite{zwitterDigitalIdentityBlockchain2020} -- that is the individual is responsible for their presence in the digital world, and this is not mediated for them by another entity.  This transition is illustrated by Bouma \cite{boumaSelfSovereignIdentityShifting2019} at Figure \ref{fig:bouma} -- whereas previously entities, such as Google or Facebook, would issue authentication for users, with the user remaining outside the circle of influence; the new model places them at its heart.

This vision was central to the artefacts delivered in this research.  The identity the individual uses to access calendar data is not provided by any third party, but rather is a cryptographic proof that only the individual possesses.  Although institutions will, rightly, play a part in deciding whether or not to grant an individual access to their data (as with \lstinline{CalAuth}); this is in the form of acknowledging the individual's identity, rather than the institution deciding what that identity is.

From the perspective of digital preservation, this change in focus has pros and cons.  The advantage in self-sovereignty is it addresses the `organisational' threat to data.  As Chapter \ref{chap:Context} identified as we move between organisations, we also have to relinquish the identities that those organisations created for us, and with it the data that belongs to those identities.  The artefacts delivered by this project demonstrate this is no longer necessarily the case.  Once you have been granted access to an organisation's data within a certain date range -- unless the organisation specifically forbids you -- then you will continue to have access to that data regardless of your association with that organisation.  Indeed the organisation could cease to exist, yet your access to that data, would live on unaffected.  This is a starkly different model to the current norm.

But there are disadvantages too -- making an individual `sovereign' comes not only with rights, but also responsibilities.  The primary responsibility, from the perspective of digital preservation, is that they must safeguard the means of accessing their data -- namely their key.  Although superficially one might imagine that users would be incentivised not to lose their keys as they are, with many DLs, associated with financial value -- the literature suggests mistakes will occur.  Krombholz's study of Bitcoin users \cite{krombholzOtherSideCoin2017} found a fifth of participants had irretrievably lost digital assets through accident or error.  Indeed this is nothing new; Whitten and Tygar \cite{whittenWhyJohnnyCan1999} demonstrate end users have been struggling with cryptography since at least 1999.

Although many of these problems require a solution at an infrastructure or ecosystem level, an individual developer will still need to understand a range of possibilities when selecting a platform.  Some potential mitigations are:

\begin{itemize}
	\item \textbf{Biometrics}. One method to prevent users from losing their keys is to integrate them with biometric data, e.g. fingerprints or retinal scans.  Cryptocurrency wallets offering biometric authentication already exist \cite{phillipsCENTBiometricWallet2020}.  Although the current state of technology uses biometrics instead of a memorised password, which then provides access to cryptographic keys; research is being conducted into using biometrics themselves to generate a key \cite{changBiometricsbasedCryptographicKey2004} meaning it is not tied to any one device. Of course this also has drawbacks: depending on the mechanism, keys could still be lost through bodily injury, obtained through physical force or in the event of death lost irretrievably.
	\item \textbf{Multi-signature}.  Cryptographic keys can also be managed using multi-signature arrangements.  Here a key is `split' between several people who cannot access data individually, but can access it when using their keys in concert.  For instance Alice might decide that as a back-up to her biometric retinal-scan key, she also creates a multi-signature key shared by a close family member and her lawyer with instructions only to use in case of incapacity or death.  Commercial solutions offering multi-signature options for Bitcoin are already available \cite{lylesCasaHighSecurity2020}.  Depending on the adoption of DLT it is plausible for public authorities to conduct a role; the General Register Office for instance might have a multi-signature capacity in ensuring assets are transferred after death.
\end{itemize}

One additional option for the developer is to use a `permissioned' rather than `permissionless' DL as discussed at Section \ref{sec:Overview}.  In this instance, key loss can be remedied by those who grant permission to this system.  Of course this also devalues the whole concept of `self-sovereignty' so it can not really be considered a solution, more of a side step. 
			
Clearly these issues are complicated, and the field has still not developed user-friendly responses to the problem that `self-sovereignty' raises.  However issues of control and identity are likely to feature in the majority of DApps, which suggests the following guidelines:

\begin{enumerate}
	\item \textbf{DApp identifies users}.  A basic requirement for control and identity.
	\item \textbf{DApp encrypts personal data}.  The default position of the Ethereum DL is that all data can be read, which is a drawback for DApps dealing with personal data.
	\item \textbf{DApp permits transfer of ownership}.  A DApp that is designed for the long term, will likely require mechanisms to permit flexibility in the relationship between the digital object and the community it serves (see continuum approach at Section \ref{sec:Threat}).
	\item \textbf{DApp implements Role Based Access Control}.  Again this takes up the notion of the continuum approach, but casting it wider realising many parties may need different forms of access to the DApp.
	\item \textbf{DL / DApp implements user-friendly self-sovereignty}.  The question of a user safeguarding their keys (and the succession of keys after death) is not an easy one, but one that must be considered for digital preservation.
\end{enumerate}

\subsection{Usability}

The final guideline concerns usability, or what might be called non-functional requirements \cite{glinzNonFunctionalRequirements2007}; a system might do everything required of it as regards accessibility, control, etc; but still be excessively difficult for a community to use and hence not be.

This is a particular risk with DLT, and one can argue has slowed its adoption to date \cite{gurgucCryptocurrenciesOvercomingBarriers2018}.  It is not a simple activity for a user to interact with a DL -- typically the interface is not natively included within most applications.  The majority of browsers require an extension (e.g. Metamask) to interact with a DL -- and even when that is installed it requires a whole new vocabulary and set of practices: `seeds' have to be carefully recorded, `gas' has to be accounted for and the addresses used to interact with both contracts and users are not human readable.  Furthermore even once that is accounted for typically the speed of transactions are far slower than those we expect in today's always-on world.

That is beginning to change.  Browsers with native DLT integration are now available \cite{bondyRoadBrave2019}, human readable names are beginning to replace unreadable addresses \cite{johnsonEthereumNameService2020} and increasing transaction time is a core feature of a number of projects \cite{bairdHederaPublicHashgraph2019}. 

The takeaway for developers with this guideline is they need to be mindful of the usability difficulties, which are numerous, that users will face; and where possible try and choose systems which minimise this.

\begin{enumerate}
	\item \textbf{User can complete use case}.  The DApp has to fulfil a minimal set of requirements.
	\item \textbf{DApp is performant}.  DApps have to be comparable to their centralised equivalents in terms of performance, although a difference is acceptable, this has to be tolerable.  This is most noticeable in terms of transaction speed, but applies equally to requirements such as reliability, etc.  
	\item \textbf{DApp scales for multiple users}.  Again any smart contract has to be scale to the level required.
	\item \textbf{DApp uses human readable attributes}.  Currently many DL related attributes are not human readable (e.g. addresses), which prevents usability.
	\item \textbf{DApp uses native interface}.  DLs integration is currently not built into many applications -- e.g. there is no calendar application that can natively connect to a DL.  Until applications are able to natively interact with DLs, keeping much of the authentication process in the background, usability will suffer.
\end{enumerate}

\section{Evaluation: Deliverable 5}\label{sec:DiscEval}

Having provided an overview of the guidelines (Table \ref{table:guideline_criteria}) relevant to designers of DApps looking to further digital preservation; an assessment of the Ethereum Calendar DApp against the contexts provided by Alice (Engineer) and Bob (Psychologist) at Chapter \ref{chap:Context} is shown at Table \ref{table:evaluation}.  The DApp is assessed to have scored a total of 7 out of a maximum value of 25.  How this score was arrived at is discussed below.

				\begin{table}
				\centering
				\begin{tabular}{ l l  } 
					\toprule  
					Criteria & Score \\
					\midrule 
					Storage & 0 \\ 
					Accessibility & 3\\ 
					Integrity & 1 \\ 
					Control \& Identity & 1 \\ 
					Usability & 2 \\ 
					\textbf{Total} & \textbf{7} \\ 
					\bottomrule
				\end{tabular}
				\caption[Evaluation of Ethereum Calendar against guidelines]{Evaluation of Ethereum Calendar against guidelines (author's own work)}
				\label{table:evaluation}
			\end{table}

\subsection{Storage}\label{sec:EvalStorage}

\textbf{Score: 0}

As discussed at Section~\ref{sec:storage} due to the limitations of storage within the Ethereum DL the \lstinline{CalStore} contract would quickly become unusable due to memory constraints; even if the current code was refactored to introduce greater efficiency.  The DApp therefore fails to achieve even the initial `solution has capacity for maximal use case' (1).  The DApp does use a common standard to store data (RFC 5545) - but this is not counted due to its failure to achieve the minimum score.

A more permanent solution would involve a third-layer solution (see Section \ref{sec:IntroSummary}) as has been demonstrated in previous research \cite{zichichiEfficiencyDecentralizedFile2020, grabisBlockchainEnabledDistributed2020}.  This is therefore more a reflection of implementation than a systematic problem -- although it is noted that third layer DDbs are themselves relatively immature. 

Given this low score the DApp would not serve the needs of either Alice nor Bob on this criteria.

\subsection{Accessibility}

\textbf{Score: 3}

The DApp scores, relatively, well here due to choice of platform: Ethereum.  By DL standards this is an active platform (1), with an open source ecosystem (2) and an active research community (3).  However like the state of all DL at the moment it is difficult to positively point to real world use cases (outside of speculation) with corporates (4) or governments (5).

At this point it is worth reflecting on exactly how the DApp leverages the accessibility offered by the platform to further digital preservation.  The artefact has two `presentation' layers: it can be accessed via a web interface (which allows the user to write to the DL) or via an API which provides an iCalendar feed read by calendar clients (e.g. MS Outlook).  By definition both of these presentation layers are centralised -- both web site and API are served by cloud provider Vercel \cite{vercelDevelopPreviewShip2020}.  Were this site to go offline then access would cease -- however the data would remain preserved in the DL.  As the source code for the DApp will be made publicly available \cite{hackmanForgetmeblockEthereumCalendar2020}; any party could construct whatever presentation layer interface they choose and restore access.  As \lstinline{CalAuth} and \lstinline{CalStore} use the same interface (as demonstrated at Figure \ref{fig:eth-cal-sequence}) the presentation layer is ambivalent as to whether it is connecting to the storage itself or an authenticator (or indeed even an authenticator coupled to another authenticator).  It is therefore truly accessible -- as long as the developer understands the open-source interface to the DL a connection can be established.

Of course the interface at the moment is purely to enable this research -- however were the concept of a DL hosted calendar to enjoy community support then this could be formalised into an Ethereum Request for Comments (ERC).  ERC's are used to standardise interfaces to Ethereum data constructs, possibly the best known is the ERC20: a token that can be exchanged for Ether and transferred between users to represent other assets.  The adoption of an ERC would mean a standard mechanism for communicating with an Ethereum based calendar could be introduced into a range of different software (e.g. desktop applications, smart phone applications, etc).  

Assuming a wider level of adoption the design of the artefact goes a considerable way to ensuring digital preservation.  The data within \lstinline{CalStore}, to some extent breaks free of the three pillars of threat (see Section \ref{sec:Threat}):  it is contained on many different forms of hardware, it is encoded in a format that is well understood and is likely to be so for some time (i.e. RFC 5545) and it is less at risk of organisational change.  That last factor is the most pivotal: once you have been given access by an institution via \lstinline{CalAuth} then it does not matter what happens to that organisation -- as long as they do not amend the contract you continue to have access to the data.

This element of evaluation therefore shows both how the artefact itself might fall short of a solution, but it leads the way for future approaches.  The DApp would partly serve the needs of Alice and Bob in this regard.  Both actively change career paths and this solution would allow both to access their data as they progress from one stage of their life to another.  One event in particular brings this into focus: when Alice's Bath consultancy goes into insolvency, an outsourced centralised calendar provider would almost certainly have meant her record of who she met when would have disappeared with the company; unless Alice had the unusual foresight and technical literacy to transfer her records.  Using a \lstinline{CalAuth} like solution would ensure her access to her records remained uninterrupted regardless of the fortunes of the company.

\subsection{Integrity}\label{sec:EvalIntegrity}

\textbf{Score: 1}

A flexible score of 1 has been awarded in this category even though by strict definitions this is not merited.  Testing did take place, however it fell short of 80\% coverage, primarily due to reasons of time pressure exacerbated by the Coronavirus pandemic.  However this is somewhat compensated for (and hence 1 has been awarded) by extensive use of industry standard libraries as listed:

\begin{itemize}
	\item \textbf{OpenZeppelin's AccessControl} and \textbf{Ownable} \cite{openzeppelinAccessControl2020}.   OpenZeppelin are a project whose libraries are ``reliable building blocks'' \cite{olivaExploratoryStudySmart2020} and are used in the literature as a byword for responsible blockchain development \cite{article}.
	\item \textbf{BokkyPooBah's DateTime Library} \cite{theofficiousbokkypoobahBokkyPooBahsDateTimeLibrary2020}.  Although not as reputable as OpenZeppelin this library has 99 stars on GitHub, has been included within high profile projects \cite{ruizPolymathBokkyPooBahsDateTimeLibrarySol2019} and is acknowledged within the community \cite{ariasListSolidityLibraries2020}.
\end{itemize}

If sufficient time for testing had been available however the project would still not have reached a score of 3, because although subjective, it is deemed that the Ethereum DL is not `sufficiently mature and stable.'  Ethereum 2.0 is imminent and although it is likely this will have a limited effect on DApps already deployed \cite{buterinEth1Eth2Transition2019}, this is not guaranteed so impinging on digital preservation.  As no funding was available for this project contracts were not audited, nor was formal verification investigated.

Were this DApp to progress to a production ready state, consideration would have to be given to the amount of computation conducted `on-chain' (see Section \ref{sec:CalStore}) -- particularly in the converting of Unix timestamps to human readable dates.  A production version of the DApp would likely simply produce a RFC 7265 (jCal) output and convert as required within the presentation layer.  Not only would this reduce gas costs, the lower code complexity would likely reduce the likelihood of errors.

It is unlikely this score would satisfy the needs of either Alice nor Bob.  Both work in industries with strong requirements for data integrity.  Bob's clinical role would place great emphasis on date and times of appointments with patients for instance, and any health provider would want to ensure a high degree of accuracy.  Errors in the code might mean appointments are not stored correctly, notified or shared more widely than designed.  Neither industry would likely be content in placing one of their major business processes on a platform not deemed sufficiently mature.  A production ready version of this DApp would therefore necessitate a higher score for both participants. 

\subsection{Control \& Identity}\label{sec:EvalControl}

\textbf{Score: 1}

\begin{figure}
				\centering
				\includegraphics[width=1.0\columnwidth]{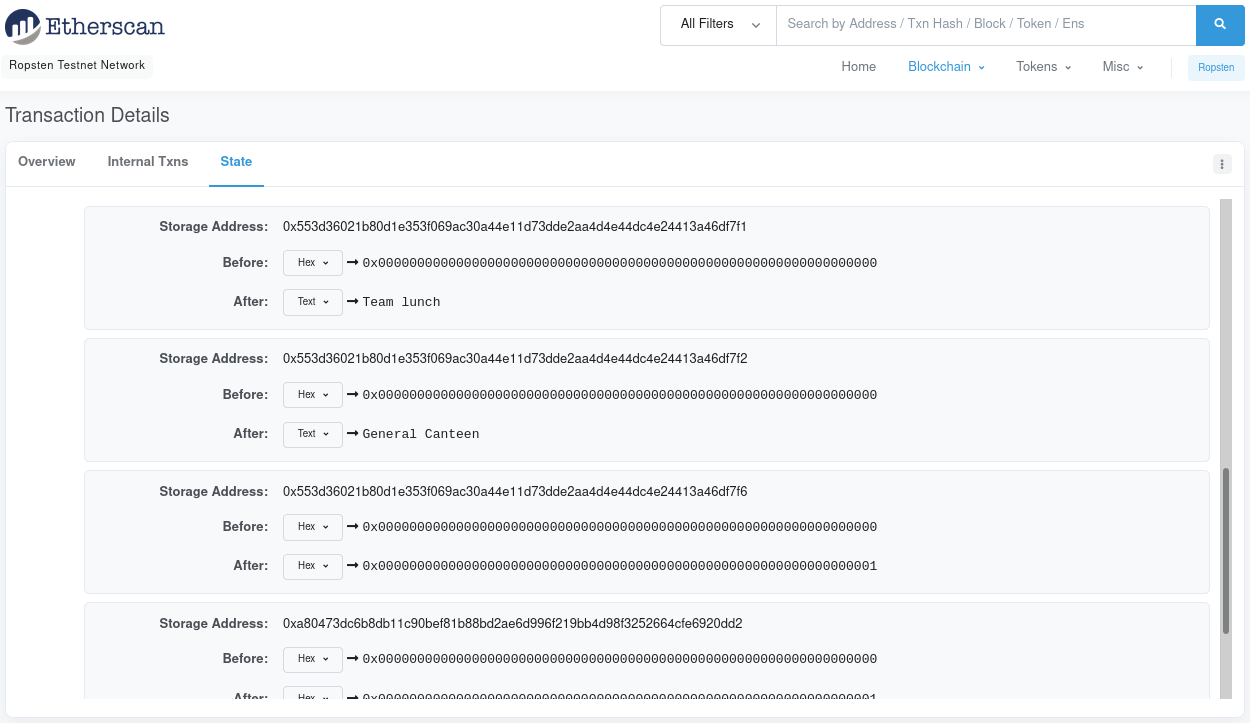}
				\caption[Ropsten transaction hash details]{Ropsten transaction hash details \cite{etherscanRopstenTransactionHash2020}}\label{fig:decode}
			\end{figure}

The DApp succeeds in identifying users (1), however it fails in ensuring privacy for personal details (2) -- as transactions are not encrypted on the Ethereum DL, they can in theory be read by anyone prepared to extract it.  This is clearly illustrated at Figure \ref{fig:decode}, examining a transaction associated with the \lstinline{CalAuth} contract using an online block explorer reveals details associated with the calendar at Figure \ref{fig:ethcal-oulook}.

As well as being able to read events via a web client, the artefact was deliberately created with an externally facing API that allowed users to receive an \textit{ics} feed directly into their calendar client.  This required knowledge of the user's address, so anyone in possession of this would be able to read a user's events.  This was necessary as calendar clients (e.g. MS Outlook) are not currently able to directly access a DL, unlike Web browsers.  It would however be possible to set up a CalDAV server which interfaces with a DL that users sign into and in that way receive an authenticated feed (see Section \ref{sec:FutureResearch}).

Clearly the privacy aspect of the DApp requires improvement, however this is ultimately a question of resource rather than technical limitations as previous research has demonstrated \cite{zyskindDecentralizingPrivacyUsing2015, grabisBlockchainEnabledDistributed2020}.

Turning to the unscored aspects: the artefact permits transfers of ownership (3) via \lstinline{CalAuth} meaning that organisations themselves are able to adapt and pass on their administrative rights to their successors.  This opens up the possibility that when an organisation has reached the end of its useful life ownership could be passed to an entity solely involved in digital preservation -- e.g. the National Archives -- so that the data contained can be utilised by wider society and future researchers.  This strongly ties the software artefacts produced in this project with the continuum approach to digital objects explored in Section \ref{sec:Threat}, where the use of an object and the relationship of a community to it changes over time and space \cite{huvilaContinuumThinkingContexts2014}.

Additionally due to \lstinline{CalAuth} and the browsers integration with a DL conduit (e.g. Metamask) only authorised users are allowed to write events into the DApp.  Even were a user to gain command line access to the DApp, without the private key or the necessary permissions, they would not be able to write.  Therefore role based access control (4) is achieved.  Again this ties into the continuum aspect - the relationship of an individual and a community to a digital object constantly changes and this functionality allows a more fine grained approach.  Although the DApp implemented a relatively simple read-only versus read-write date-ranged permissions schema, this acts to demonstrate what could be a far more intricate arrangement.  The final score (5) regards implementing a `user friendly self-sovereignty' is not attempted and represents a considerable body of work in its own right for the community as a whole. 

From the perspective of the scenarios a lack of privacy would be a critical flaw for both Alice and Bob. Bob would have patient confidentiality as a primary concern.  Alice likewise may have operational security requirement from her period in the Royal Navy which would require encryption.  Regardless of these specific use cases privacy is an essential component of PIM, even for those used to a world of social sharing \cite{ozdemirAntecedentsOutcomesInformation2017}.

\subsection{Usability}

\textbf{Score: 2}

It was deemed that the user could complete the use case (1), even though that was a minimal, not feature-rich one.  The DApp was also relevantly performant (2) -- as the web presentation layer cached events in browser the user did not have to wait for transactions being read onto the DL before being able to view.  The API, which allowed the user to view events within a regular calendar application, was comparable in terms of speed to a centralised alternative.  Due to time constraints the DApp was never tested for multiple users (3) -- and indeed due to storage limitations (see Section \ref{sec:EvalStorage}) it would not have had enough memory for multiple users.  Likewise it suffered from many of the problems that DApps routinely suffer from -- non-human readable attributes (4) and not using a native interface (5).

There may be some divergence here dependent on Alice and Bob's particular scenarios.  Due to Alice's occupation as a Marine Engineer we can expect considerable parts of her career to be spent at sea.  The artefact in its current form -- and indeed most current DL solutions -- require a reliable connection to the internet.  Clearly the DApp would become unusable if a connection could not be established and events could not be saved or recalled.  There are options though for rectifying this.  In the shorter term, given correct access levels, a cache could be implemented onboard to save data until reconciliation was achieved.  In the longer term, depending on the adoption of DLT technology, a DL node onboard could be feasible, in much the same way that a ship may have servers now.  Clearly however there are considerable architectural challenges to overcome before that becomes a possibility.  It is possible therefore within Alice's context the DApp scores 0 as it cannot sustain basic usability.

Bob alternatively, who spends his life in good connectivity, mostly within academia, government and the health service; is different.  The artefact produced can be considered at an alpha stage, from a usability perspective it could serve as a minimum viable product.

\section{Summary}

The DApp gained a total score of 7 out of a maximum of 25.  This partly reflected the nature of developing a design science artefact under social distancing conditions imposed by the Coronavirus pandemic, rather than production ready software.  But it also reflected the realities of designing a DApp for digital preservation at this point in time.  The DL ecosystem is immature and not enough real world business processes exist as DApps for an assessment that this technology will stand the test of time - from this perspective it is judged that the guidelines worked well.

Something not fully discussed within the evaluation is the actual maintenance of the contracts themselves.  This is a complex topic, but ultimately, software is rarely born fully formed and error free and even if it is then the environment it depends on will  undoubtedly change~\cite{hongWhyGoodSoftware2008}.  If it is unrealistic to expect \lstinline{CalAuth} and \lstinline{CalStore} to be frozen in aspic -- who then is responsible for their maintenance: assuming they are even hosted on a DL that allows this to happen?  What are the incentive structures that would allow this to happen and the governance arrangements to ensure agreement?  Of course it is impossible to predict -- but there are comparisons.  The community that grew up and sustained Linux has proven successful despite starting from humble beginnings and without obvious financial incentives, Carvalho et al suggest a similar path for DLT \cite{carvalhoWhatHistoryLinux2020}.  The artefact as designed by this research proposed one \lstinline{CalStore} instance, however were an ERC standard to be devised there could be multiple contracts offering this service.  At the same time these could all be easily and automatically identified -- in the same way that all ERC20 tokens on the Ethereum mainnet can be identified now \cite{etherscanTokenTracker2020}.  Theoretically then \lstinline{CalStore} contracts could be offered by a variety of entities offering different value propositions, but within a standard framework.  Some of these entities could also group together, for instance professional groupings may want some derivation of \lstinline{CalStore} or \lstinline{CalAuth} for their own ends, but still within the general standard proposed by an ERC.  Indeed going even further, there are efforts to link DLs together \cite{kanMultipleBlockchainsArchitecture2018}, theoretically meaning instantiations of \lstinline{CalStore} and \lstinline{CalAuth} could sit on different DLs, but still communicate to users on other networks.  Clearly though this is some way ahead of the current state of technology.

This is a large problem space -- digital preservation concerns planning for the future, so by definition revolves around unknowns.  There is not a `solution' as such; and indeed the artefacts this project has produced, as the evaluation shows, do not adequately perform against the same guidelines that the project demands.  Indeed taking a philosophical approach one might argue that digital preservation is inherently unsolvable -- when one factors in a long enough timeline the second law of thermodynamics suggests that entropy will eventually win \cite{seifertCanWeDecrease1961}.

From a less philosophical position however, it is likely that only time will tell whether DLT; either Ethereum or a different technology; offer valid mechanisms for promoting digital preservation.  What the artefacts do however is paint a scenario of an imagined world where decentralised systems free us from some of the problems currently implicit in ensuring digital preservation.


\let\textcircled=\pgftextcircled
\chapter{Conclusions}\label{chap:Conclusions}

\initial{T}his project following a review of the literature has produced a number of deliverables in the form of evaluated software artefacts and guidelines.  The software artefacts implemented a RFC 5545 compliant calendar smart contract hosted on the Ethereum DL which could be accessed both via a web interface and through a calendar client (e.g. MS Outlook).  Furthermore role based access control, using an industry standard library, was applied to this calendar via an interfacing smart contract so simulating institutional like control.  The presentation layer (e.g. web or desktop application) could be used either directly into the calendar or via the authenticator.

Guidelines, formulated both by reflection and a study of the literature, aimed at developers and researchers designing DApps with digital preservation in mind were also produced.  The software artefacts developed were then evaluated against these guidelines using as context scenarios featuring two professionals who encountered typical digital preservation challenges over the course of their careers.  The Ethereum Calendar DApp scored a total of 7, out of a maximum of 25, when evaluated against the project's guidelines.  This suggested that in current form it would not serve digital preservation well.  The guidelines produced by this project were deemed to have been validated by their robust assessment of the DApp.

\section{Research contribution}\label{sec:ConcContribution}

This project contributes to research in the following way:

\begin{itemize} 
	\item Presenting the first research to date that applies DLT to the field of PIM (as defined at Section \ref{sec:pim}) and furthering design science state of the art by a novel implementation of a calendar application on the Ethereum blockchain.
	\item Extending current research in utilising DLT in digital preservation, namely by enacting a continuum approach within a DL that allows for transfer of ownership of digital objects as they transition from individual to collective relevance.
	\item Providing guidelines for future use of DLT within digital preservation; and demonstrating the utility of these guidelines via an evaluation of this project's software deliverables.
\end{itemize}
	In accordance with design science principles, post-submission, all source code will be made available on a public GitHub repository with an opensource licence, the software artefacts will be made live online and a blog post will be published linking code, live artefacts and implementation details \cite{hackmanForgetmeblockEthereumCalendar2020}.

\section{Limitations}

As the evaluation showed the artefacts scored poorly against the digital preservation guidelines produced by this research.  This reflected partly the time constraints of the project; but also a realistic view of the utility of DLT for digital preservation at this point of time.  The two critical flaws regarded storage and control.  A design choice was made to store persistent data on the Ethereum DL, rather than a decentralised third layer (see Section \ref{sec:IntroSummary}) -- this was very limiting as the amount of data that can be stored in a smart contract is insufficient for this use case.  The second critical flaw concerned control or privacy -- as the data was not encrypted it was viewable to anyone that was prepared to interrogate the blockchain.

One might argue that the choice of DL, Ethereum, was itself a limitation.  Ethereum is relatively slow in conducting transactions and its programming language (Solidity) has been criticised for making it difficult to write error free smart contracts.  However given the novelty of the DLT landscape there was no valid option that would have lived up to the guideline's requirement for maturity and stability of platform, while at the same time allowing the flexibility to implement a PIM application.

Outside of technical issues, the mode of evaluation used in this study was by comparing guidelines and deliverable against example scenarios.  This decision was taken as the software artefacts were ultimately proving a concept, so testing them with actual users would have likely concentrated on the user experience (e.g. interface), which would have had little merit.  It is recognised however that external participation in the research may have produced more rigorous evaluation and feedback.

\section{Future Research and Development}\label{sec:FutureResearch}

Looking first to the software artefacts themselves there are several directions in which these could be improved -- primarily by addressing the limitations above.  A third layer decentralised database; e.g. IPFS \cite{benetIPFSConentAddressed2014}, BigchainDB \cite{bigchaindbBigchainDBBlockchainDatabase2018}, etc; would allow persistent data to be stored outside of the smart contract itself so allowing its enduring operation.  Similarly the artefact could be designed to encrypt data, which would solve control issues.  Consideration would also have to be given to what DL to actually use -- Ethereum was primarily chosen for ease of development; but a better resourced effort might consider using a hashgraph such as Hedera \cite{bairdHederaPublicHashgraph2019}, which may have improved transaction speeds.

There was considerable discussion in this project as to the lack of applications with `native' connections to DLs.  To perform write operations on the DL a browser extension, Metamask, had to be utilised.  If DLT is going to succeed in its ambition to become a ``world computer'' \cite[p.~25]{antonopoulosMasteringEthereum2018} as opposed to its emphasis on buying and selling tokens, then the interface between DL and other internet services will have to be strengthened.  One fascinating approach that would take this work further would be to investigate DL integration within a CalDAV server \cite{dusseaultCalendaringExtensionsRFC2007} -- in practical terms this would allow a read-write interface between calendar clients and a DL; as opposed to the read-only API in this research.  An alternative route to the same functionality would be to write a Metamask-like extension for a desktop calendar application (e.g. Mozilla Thunderbird) which allowed it to directly query a DL.  This though, unlike a CalDAV server, would need to be re-written to work with different desktop applications; and proprietary applications may not allow integration.

Looking at PIM specifically there are a number of other applications that could be examined within a DLT environment, for instance a contacts manager, to-do list, etc.  Designing these applications would likely prove further valuable learning points in the utility of building on DLT.  

To conclude, although DLT is still immature, it has considerable and exciting potential.  The field is therefore ripe for research exploring real world DApps in digital preservation; PIM and otherwise.  This presents a genuinely new way to go beyond the theoretical and instead employ practical strategies to bring about ``the end of forgetting'' \cite{rosenWebMeansEnd2010}.

\clearemptydoublepage

%
\appendix
\chapter{Message Time Store Source Code}
\label{app:msg-time-store}
The Solidity source code for Deliverable 1, \textit{Message Time Store}, is shown below.  This also illustrates comments that conform to the Ethereum Natural Language Specification Format (NatSpec); this project's digital preservation guidelines highlight using standards where possible and thorough code documentation (see Section \ref{sec:Integrity}).\\

\begin{lstlisting}[language=Solidity,caption={TimeStore.sol},captionpos=b,label={lst:msgtimestore}]
// SPDX-License-Identifier: Unlicense
pragma solidity ^0.6.1;
pragma experimental ABIEncoderV2;

/// @title Message Time Store
/// @author James D Hackman: preciouschicken.com
/// @notice Stores a string from address, releases string after specified time
contract TimeStore  {

    struct StoredData {
        uint id;
        string storedMsg;
        uint unlockTime;
    }
    StoredData[] private userStoredData;
    mapping(address => StoredData[]) private store;

    /// @notice Stores incoming messages from address 
    /** @dev nextId starts at 1, so zeros can be removed by client.
      Neccessary as memory array in getMsgTimed can return empty arrays.
      */
    /// @param _storedMsg string to be stored
    /// @param _unlockTime unix time for _storedMsg to be stored until
    function storeMsg(string memory _storedMsg, uint _unlockTime) public {
        StoredData[] memory currentData = store[msg.sender];
        uint nextId = currentData.length + 1;
        StoredData memory newData = StoredData(nextId, _storedMsg, _unlockTime);
        userStoredData.push(newData);
        store[msg.sender] = userStoredData;
    }

    /// @notice Returns unlocked messages, as array of objects
    /// @dev For any message not unlocked returns an object with blank values
    /// @return StoredData[] array of StoredData struct 
    function getMsgTimed() public view returns (StoredData[] memory) {
        StoredData[] memory currentData = store[msg.sender];
        StoredData[] memory unlockedData = new StoredData[](currentData.length);
        uint unlockedDataIndex = 0;
        for (uint i = 0; i < currentData.length; i++) {
            if (currentData[i].unlockTime <= now) {
                StoredData memory newData = StoredData(
                    currentData[i].id, 
                    currentData[i].storedMsg, 
                    currentData[i].unlockTime);
                unlockedData[unlockedDataIndex] = newData;
                unlockedDataIndex++;
            }
        }
        return unlockedData;
    }

}
\end{lstlisting}

\clearemptydoublepage

%
\backmatter 
\refstepcounter{chapter}

			\chapter{Bibliography}
			\printbibliography[heading=none]

\end{document}